\documentclass[traditabstract]{aa} 
\usepackage{graphicx}
\usepackage{txfonts}
\usepackage{natbib}
\newcommand{\CIV}{{\rm C~{\sc iv}}} 
\newcommand{\CIII}{{\rm C~{\sc iii]}}}

\newcommand{\cm}{{\rm cm}}

\newcommand{\kms}{{\rm km}\,{\rm s}^{-1}}

\newcommand{\lya}{Lyman-$\alpha$}
\newcommand{\lyb}{Lyman-$\beta$}

\def\Sec#1{Section~\ref{s:#1}}
\def\Eq#1{Eq.~\ref{eq:#1}}
\def\Fig#1{Fig.~\ref{fig:#1}}
\def\Tab#1{Table~\ref{t:#1}}

\begin{document}
   \title{A Principal Component Analysis of quasar UV spectra at $z\sim 3$}
   \titlerunning{PCA of  QSO UV spectrum at $z \sim 3$}

   \author{ I. P\^aris 
          \inst{1}
          \and
          P. Petitjean 
          \inst{1}
          \and
          E. Rollinde 
          \inst{1}
          \and
          E. Aubourg 
          \inst{2}
          \and
          N. Busca 
          \inst{2}
          \and
          R. Charlassier 
          \inst{3}
          \and
          T. Delubac 
          \inst{3}
          \and
          J.-Ch. Hamilton 
          \inst{2}
          \and
          J.-M. Le Goff 
          \inst{3}
          \and
          N. Palanque-Delabrouille 
          \inst{3}
          \and
          S. Peirani 
          \inst{1}
          \and
          Ch. Pichon
          \inst{1}
          \and
          J. Rich 
          \inst{3}
          \and
          M. Vargas-Maga\~na 
          \inst{2}
          \and
          Ch. Y\`eche 
          \inst{3}
                    }

   \institute{$^1$~UPMC Univ Paris06, Institut d'Astrophysique de Paris, UMR7095-CNRS, 
	F-75014, Paris, France\\
              \email{paris@iap.fr}\\
   $^2$~APC, 10 rue Alice Domon et L\'eonie Duquet, F-75205 Paris Cedex 13, France\\
   $^3$~CEA, Centre de Saclay, Irfu/SPP, F-91191 Gif-sur-Yvette, France}
	\authorrunning{P\^aris et al.}

   \date{Received \today \ / Accepted }

  \abstract{
From a Principal Component Analysis (PCA) of 78 $z\sim3$ high quality quasar spectra in the SDSS-DR7, we derive the principal components
characterizing the QSO continuum over the full wavelength range available. 
The shape of the mean continuum, 
is similar to that measured at low-$z$ ($z\sim 1$), but the equivalent width of the emission lines
are larger at low redshift.
We calculate the correlation between fluxes at different wavelengths and find 
that the emission line fluxes in the red part of the spectrum are correlated with that in the blue part. 
We construct a projection matrix to predict the continuum in the \lya\ forest from the red part of the spectrum.
We apply this matrix to quasars in the SDSS-DR7 to derive the evolution with redshift of the mean flux in the \lya\ forest due to the absorption by the intergalactic neutral hydrogen.
A change in the evolution of the mean flux is apparent around $z\sim 3$  in the sense of a steeper decrease of 
the mean flux at higher redshifts. 
The same evolution is found when the continuum is estimated from the extrapolation of a power-law continuum fitted
in the red part of the quasar spectrum if a correction, derived from simple simulations, is applied.
Our findings are consistent with previous determinations using high spectral resolution data. 
We provide the PCA eigenvectors over the wavelength range 1020$-$2000 \AA\ and the distribution of their weights 
that can be used  to simulate  QSO mock spectra. 
}
   \keywords{Methods: numerical --- galaxies: intergalactic medium, quasars
               }
   \maketitle
   
%

\section{Introduction}

At high redshift, most of the baryons are located in the intergalactic medium \citep[IGM; {\em e.g.}][]{petitjean93} 
where they are highly ionized by the UV-background produced by galaxies and QSOs \citep{GunnandPeterson}, at least 
since $z\sim 6$ \citep[]{Fan06, Becker07}. 
The large absorption cross section of the H~{\sc i} Lyman-$\alpha$ transition
implies that the small fraction of neutral hydrogen in the IGM produces the so-called Lyman-$\alpha$ forest composed of 
numerous absorption lines detected in the spectra of high redshift quasars 
\citep[see][ for a review]{Lynds,Rauch98}.

Analytical models \cite[]{Bi92} and numerical $N$-body simulations 
\citep[]{Cen94,petitjean95,Zhang95, Hernquist96, Theuns98, riediger98}
have been very successful at reproducing the properties of the
Lyman-$\alpha$ forest as measured from high spectral resolution, high SNR data obtained 
with the Ultraviolet and Visual Echelle Spectrograph (UVES) on the Very Large Telescope  
\citep[VLT, {\em e.g.}][]{Bergeron04,Kim07} and {\sc HIRES} on the Keck telescope \citep[{ {\em e.g.}}][]{Hu95}.
The overwhole picture tells that lower column-density H~{\sc i} absorption lines trace the filaments of
the \lq cosmic web\rq, and higher column-density absorption lines trace
the surroundings of galaxies.
Detailed studies of absorption line properties and of their clustering properties 
along one or several adjacent lines of sight give additional constraints on the ionization history, correlation length, matter 
power spectrum etc... \citep[see {\em e.g.}][]{petitjean98,Croft98, McDonaldal05a,theunsal06}.
The next generation of quasar surveys, from BOSS \citep[SDSS-III,][]{Schlegelal07,Eisenstein11} 
to Big-BOSS \citep{BigBOSS} should provide the first  detection of Baryonic Acoustic Oscillations in the IGM at 
$z\sim 2-3$ \citep[]{slosaral09,whiteal09}.

An important quantity to measure in a quasar spectrum is the mean amount of absorption in the Lyman-$\alpha$ forest, 
$D_{\rm A}$, defined as: $D_{\rm A}~= ~1-$$<$$F$$>$ \citep{Okeal82}, where $F$ is the quasar normalized flux, 
$F=F_{\rm obs}/F_{\rm cont}$, $F_{\rm obs}$ is the observed flux and $F_{\rm cont}$ is the estimated unabsorbed continuum flux.
The absorption can be defined as well by the mean effective optical depth, $\tau_{\rm eff}~=~- \ln$$<$$F$$>$.
These quantities are sensitive to the physical properties of the IGM and have 
been used to constrain $\Omega_{\rm b}$ \citep{Rauch98,Tytleral04}, the ionization history 
\citep{Rauchal97,Kirkman2005,Bolton05,Boltonal07,Prochaskaal09} and in particular the He~{\sc ii} reionization 
\citep{bernardial2003,Theuns02c}. 
The latter could possibly induce a dip in the evolution with redshift of $\tau_{\rm eff}$ at $z\sim 3.2$
\citep{Schaye00}. 

\cite{bernardial2003} first discovered such a dip in the evolution of  the effective optical depth in the \lya\ forest using SDSS spectra. 
The existence of a feature has then been confirmed from high-resolution studies \citep{FG2008,Dallaglioal08}, 
at a more modest statistical significance but at a coincident redshift.

Intermediate resolution data have been used as well \citep{McDonaldal05a,Dallaglioal09} and
the feature is not detected. However, the methods used may not be totally appropriate and 
in the present paper, we come back to this point.
Note that \cite{FG2008} cautioned that (i) the dip interpretation is only valid if one insists on fitting a single power law to the 
background evolution, and there is no clear physical motivation for doing so and (ii)
this feature is not necessarily due to He~{\sc ii} reionization and other interpretations are also possible.

The definition of the unabsorbed quasar continuum over the Lyman-$\alpha$ forest is a critical issue \citep[e.g.][]{Tytleral04,Kim07}.
For low resolution spectra, most analysis define first a continuum redwards to the QSO Lyman-$\alpha$ emission, where there are 
only few absorption lines, and extrapolate the shape of the continuum in the Lyman-$\alpha$ forest region (see \Sec{cont-est} for 
more details). It is most commonly assumed that the QSO continuum in regions where there is no emission line is a power-law that can 
be extrapolated easily. However, 
this assumption usually neglects weak emission lines both in the red and, more importantly, within the Lyman-$\alpha$ forest region.
Because of this, \citet[][S05]{Suzukial05} have applied a Principal Component Analysis (PCA) to HST spectra of quasars at $z \leq 1$.
Quasar continua are described with a limited set of eigenvectors and a controlled sample is used to define a  
projection matrix allowing to recover the continuum in the Lyman-$\alpha$ forest from the shape 
of the continuum in the red part.

The shape of quasar continuum can evolve from $z \leq 1$ to $z \sim 3$.
In such a case, a PCA at $z \leq 1$ would not give a fair representation of quasar continuum at $z\sim 2-3$. 
In this paper, after describing the procedures in \Sec{sect1}, we take advantage of the large database provided by 
SDSS-DR7 to define a large enough sample of quasars at $z \sim 3$ on which we can apply the same procedure as in S05 (\Sec{pcaz3}). 
New eigenvectors and projection matrix
are generated and then used to predict the continuum of all SDSS-DR7 spectra. 
We apply the method to the determination of the evolution with redshift of the mean flux
in the Lyman-$\alpha$ forest (\Sec{meanflux}) and discuss the significance of the bump at $z\sim 3.2-3.4$ before
drawing our conclusions in \Sec{Conclu}.

%

\section{Procedures}
\label{s:sect1}

\subsection{Different methods to estimate the QSO continuum in low resolution spectra}
\label{s:cont-est}

The methods used to estimate the continuum in low resolution spectra
can be broadly classified as below:
\begin{enumerate}
\item A direct estimate of the continuum in the Lyman-$\alpha$ region:
  \begin{itemize}
    \item Using a spline interpolation: A cubic spline is interpolated on adaptative intervals 
between observed data points in the forest to construct a local continuum. A correction is then applied to take into account 
the fact that these data points can be affected by some absorption.
\cite{,Dallaglioal08,Dallaglioal09} apply a systematic correction
which accounts for resolution effects and line blending and is estimated from idealized Monte-Carlo simulated spectra.
This approach however neglects the possibility of continuous absorption from the smooth IGM (rather than discrete absorbers)
as well as correlations from large-scale structure. At high redshift, continuous absorption can be important and cause the true continuum to be underestimated even after applying the Monte-Carlo method to high-resolution data. \cite{FG2008} developped an alternative method to correct the continuum placement using cosmological simulations and showed that this effect is actually important at the $>$10\% level at $z=4$.

    \item Taking into account the difference between the continuum and absorption wavelength dependencies \citep[e.g.][]{bernardial2003,Prochaskaal09}:
The continuum is a property of the quasar and depends only, to first order, on the restframe wavelength ($\lambda_{\rm r} = \lambda/(1+z_{\rm em})$); while the 
absorption depends on the redshift ($z_{\rm abs}=\lambda/1215.6701-1$) only. Thus, if one separates the dependencies in the flux, 
$F(\lambda_{\rm r},z)=C(\lambda_{\rm r})\exp(-\tau(z))$, both quantities can be recovered in principle. 
  \end{itemize}
  \item Using the red part of the QSO spectrum to predict the blue part,
    \begin{itemize}
      \item A power law is adjusted to the red part of the spectrum in regions free of emission 
and absorption lines (see \Sec{zqso}). The power law is then simply extrapolated over the Lyman-$\alpha$ forest wavelength range. This procedure does not 
account for the presence of weak emission lines in the Lyman-$\alpha$ forest region. 
      \item A Principal Component Analysis applied to a reference sample describes the continuum of a quasar spectrum as a linear 
combination of eigenvectors. A projection matrix is generated and used to translate the weights of the eigenvectors describing the red 
side of the spectrum into the weights of the eigenvectors for the whole spectrum.
  \end{itemize}
\end{enumerate}

In this paper, we focus on the two last methods in which information 
provided by  absorption-free regions redwards to the Lyman-$\alpha$ emission line are  used to predict the continuum in the forest.
Both methods require first to estimate the quasar emission redshift. Our procedure to estimate this redshift and 
to provide a power-law continuum is described in the next sub-section. 
We then describe in Section~2.3 the PCA method.

	\subsection{Determination of the position of the emission lines and the power-law continuum}
\label{s:zqso}

\begin{figure*}
\centering{\includegraphics[angle=-90,width=130mm]{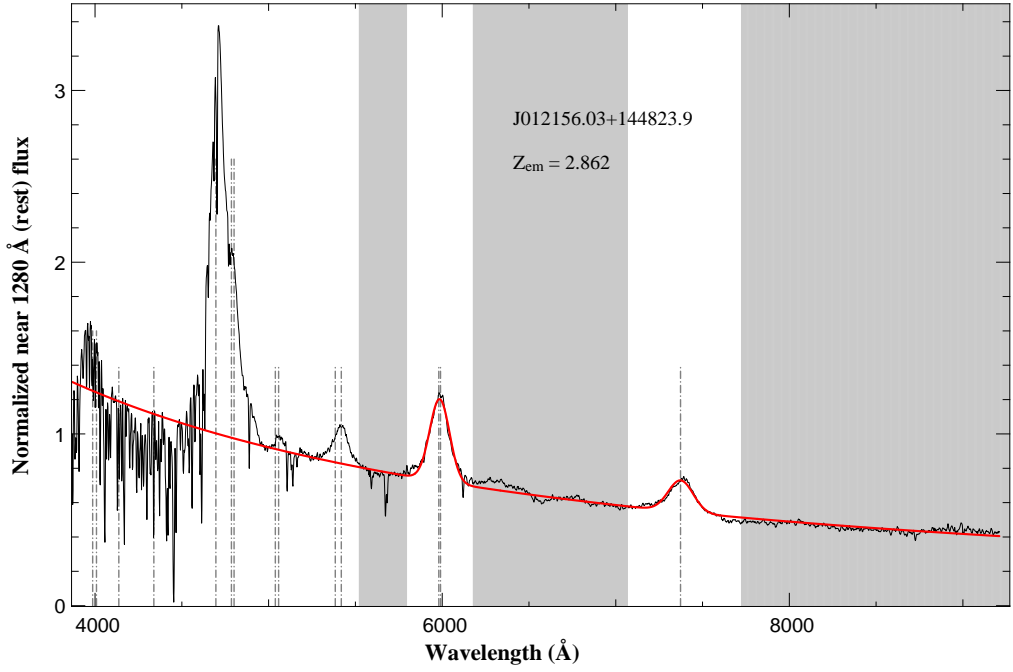}}
\caption{Illustration of the method used to estimate the emission redshift and power-law continuum of quasars 
(the quasar shown is SDSS~J012156.03+144823.9). Grey areas indicate 
the regions used to fit the power-law and the red line is the estimate of the continuum (power-law + \CIV\ and \CIII\ emission 
lines). Vertical dashed lines indicate the position of emission lines.}
\label{fig:zqso}
\end{figure*}

We derive a redshift of the quasar using C~{\sc iv} and \CIII\  emission lines in order to
be able to estimate the position of these lines and to avoid them when fitting the power-law. 
This is therefore not an attempt to derive the exact systemic quasar redshift which is known to
be shifted compared to the C~{\sc iv}-\CIII\ redshift \citep[e.g.][]{VB01,hennawi2007}.\\

Here we assume that the continuum redwards to the \lya\ emission line can be described as the sum of a power-law component and a gaussian
function for each of \CIV\ and \CIII\  emission lines. 
The redshift and the power-law component are estimated as 
follows (see \Fig{zqso} for a typical example at $z_{em}=2.862$):
\begin{enumerate}
	\item After convolving the whole spectrum with a gaussian filter of fixed FWHM~=~$250 \ \kms$, the position where the flux is maximum 
is associated to the QSO \lya\ emission line. This gives  a first rough estimate of the redshift, $z_1$.
	\item  The average of the positions of the maximum flux within a window of 100~\AA\ in the rest frame of the quasar (at $z=z_1$) around 
\CIV\ and \CIII\ emission lines provides a second redshift estimate, $z_2$. This redshift should be more accurate than $z_1$  
because the peak of the \lya\ emission is a poor estimate of the redshift due to the presence of the Lyman-$\alpha$ forest and the
blending with N~{\sc v}$\lambda$1240.

	\item A power-law component of the continuum is fitted using windows devoided of emission lines
between 1430-1500 \AA, 1600-1830 \AA\ and 2000-2500 \AA~ in the rest-frame (see grey windows in \Fig{zqso}). 
This component is subtracted from the spectrum.

	\item  Finally, \CIV\ and \CIII\ emission lines are simultaneously fitted with gaussian functions. The width and amplitude 
are independant parameters, but the two gaussian functions are bound to have the same redshift, $z_3$.
\end{enumerate}
The red line in \Fig{zqso} is the sum of the power law and the two emission lines.
The extrapolation of  the power-law component bluewards to the \lya\ emission provides a first estimate of 
the quasar continuum (red line in \Fig{zqso}).\\

\subsection{Principal Component Analysis}
\label{s:pca}

We summarize here the main steps of the method as described in \cite{francis92} and \cite{Suzukial05}.

\subsubsection{Reconstructed continuum}

A representative sample of quasar spectra at the redshift of interest must be gathered
for which it is possible to define a true continuum, $q(\lambda)$, e.g. unspoiled by
intervening absorption.  
S05 used HST spectra at $z\ \leq 1$ because the IGM is sparse at these redshift
and thus the continuum can be easily interpolated above absorption lines.
The sample of SDSS-DR7 quasars we used has a mean emission redshift of $z \sim 2.9$ and is defined in \Sec{pcaz3}.
We derived the true continuum, $q(\lambda)$, by eye
and used these fitted quasar continua in the following.

A covariance matrix {\boldmath $V$} is first calculated for the N  QSOs in the sample as:
\begin{equation}
\mbox{\boldmath $V$}(\lambda_{\rm m},\lambda_{\rm n})=
\frac{1}{N-1}\sum_{\rm i=1}^{\rm N}
\left ( q_{\rm i}(\lambda_{\rm m})-\mu(\lambda_{\rm m}) \right )
\left ( q_{\rm i}(\lambda_{\rm n})-\mu(\lambda_{\rm n}) \right ).
\end{equation}
where $\mu(\lambda)=1/N\sum_{\rm i=1}^{\rm N}q_{\rm i}(\lambda)$ is the
mean quasar continuum.

The principal components are found by decomposing the covariance matrix {\boldmath $V$} into the product of the 
orthonormal matrix {\boldmath $P$} which is composed of eigenvectors, and the diagonal 
matrix {\boldmath $\Lambda$} containing the eigenvalues:
\begin{equation}
\mbox{\boldmath $V$}=
\mbox{\boldmath $P$}^{-1}
\times
\mbox{\boldmath $\Lambda$} 
\times
\mbox{\boldmath $P$}.
\end{equation}
We call the eigenvectors (i.e. the columns of the matrix {\boldmath $P$}) the principal components, $\xi_{\rm j}$. 
The principal components are ordered according to the amount of variance in the training set they can accommodate,
such that the first principal component is the eigenvector which has the largest eigenvalue.

The distribution of the weights, $c_{\rm j}$, of the {\it j}th principal component
in \Eq{eq_reconstruction} is found from the distribution of the $c_{\rm ij}$ 
for all $i=1..N$ QSOs of the sample:
\begin{equation}
c_{\rm ij}=\int_{1020 {\rm \AA } }^{2000 {\rm \AA } }(q_{\rm i}(\lambda)-\mu(\lambda))
\; \xi_{\rm j}(\lambda)\; d\lambda .
\label{eq:eq_weight}
\end{equation}

Note that the upper limit of the integration is larger here compared to S05. This is discussed further in \Sec{pcaz3}.
A mock continuum can now be constructed using \Eq{eq_reconstruction} over the rest frame wavelength range of the spectra in the sample.
\begin{equation}
q(\lambda) \sim 
\mu(\lambda)+\sum_{\rm j=1}^{\rm m}c_{\rm j} \; \xi_{\rm j}(\lambda),
\label{eq:eq_reconstruction}
\end{equation}

\subsubsection{Predicted continuum}

The goal is to quantify the relationship between the red and blue sides of the spectra in the sample.
The first $m$ principal components $\xi_{\rm j}(\lambda)$ and their weights, $c_{\rm ij}$, are derived
as described above, using the whole rest wavelength range, 1020 to 2000~\AA.
Another set of $m$ principal components, $\zeta_{\rm j}(\lambda)$ and 
their weights, $d_{\rm ij}$, are defined using only the red rest wavelength range, 1216 to 2000~\AA. 
Finally, we solve linear equations to find a projection matrix relating $c_{\rm ij}$ and $d_{\rm ij}$.
Weights can be written in the $N \times m$ matrix form {\boldmath $C$} $= c_{\rm ij}$ 
and similarly for {\boldmath $D$}. We then use singular value decomposition techniques 
\citep{press92} to derive the $m \times m$ projection matrix {\boldmath $X$} $ = x_{\rm ij}$ 
translating weights found with the red side only into the weights for the whole spectrum:

\begin{equation}
\mbox{\boldmath $C$} = \mbox{\boldmath $D$} \cdot \mbox{\boldmath $X$}.
\end{equation}

Once matrix $X$ is known, we can estimate the continuum over the Lyman-$\alpha$ forest
for any quasar spectrum from the red part of the spectrum.
We proceed in three steps.
The weights for the red spectrum are found,
\begin{equation}
b_{\rm j}=\int_{1216 {\rm \AA } }^{2000 {\rm \AA } }(q(\lambda)-\mu(\lambda))
\;\zeta_{\rm j}(\lambda)\;d\lambda.
\label{eq:eq_weight_red}
\end{equation}

The weights from the red side $b_{\rm j}$ are translated to weights for the whole spectrum, using
\begin{equation}
a_{\rm j}=\sum_{\rm k=1}^{\rm m}b_{\rm k} \; x_{\rm kj}.
\end{equation}
 Then the continuum for the whole spectrum is built as:
\begin{equation}
p(\lambda)=\mu(\lambda)+\sum_{\rm j=1}^{\rm m}a_{\rm j} \; \xi_{\rm j}(\lambda).
\label{eq:eq_prediction}
\end{equation}

%
\section{New PCA continuum at $z \sim 3$}
\label{s:pcaz3}

The eigenvectors and coefficients as derived by S05, at low redshift, have been used to generate mock spectra in order 
to test different analysis at high redshift \citep{Dallaglioal08,Kirkman2005}. 
To our knowledge, one  attempt has also been made 
to derive a similar decomposition at high redshift using SDSS spectra \citep{McDonaldal05a}. The latter authors 
comment that this continuum determination is robust enough to infer the mean flux evolution, but unstable as far as the 
Lyman-$\alpha$ power spectrum is concerned. However the decomposition is not performed on a well controlled 
training sample as in S05 and in the present work (see below), therefore, they do not provide a projection matrix.
The analysis performed by \cite{Yipal04} is closer to our purpose. They applied a PCA to the 16,707 Sloan Digital Sky Survey DR1 
quasar spectra (0.08 $< z <$ 5.41) and reported that the spectral classification depends on redshift and luminosity. 
No compact set of eigenspectra succeeds in describing the variations observed over the whole redshift range. Besides, it seems that there is a 
differential evolution with redshift of the coefficients. Since these authors are interested in the quasar continuum only, they do not try to recover 
the exact continuum over the \lya\ forest, and consider only the observed flux (that is continuum plus absorption). \\

Due to the large difference of redshift between the quasars used in S05 and those involved in any \lya\ forest 
study using SDSS spectra, one may wonder if there is any evolution in the continuum of quasars or any change in the 
correlation between the shapes of the continuum over the forest and redwards to the \lya\ emission compared to what is found by S05. 
Deriving new components at redshift 3 should answer this question.

The other motivation for this work is to provide principal components and distributions of coefficients over a larger 
wavelength coverage than in previous studies: the S05 matrix allows to generate continua from Lyman-$\beta$  to \CIV\ emission lines 
while in the present work we will extend the wavelength coverage
beyond the \CIII\  emission line (until 2000 \AA\ in the restframe). This should in principle facilitate the extrapolation 
in the blue.

\subsection{Deriving new principal components from a sub-sample of $z\sim 3$ SDSS-DR7 quasar spectra}
\label{s:new-pca}

The difficulty of this analysis is to estimate the continuum in the \lya\ forest, where absorption can be 
neither neglected nor easily removed because of the low resolution of SDSS spectra. In particular it would be
very difficult to define the true continuum automatically.  

\begin{figure}
	\centering{\includegraphics[angle=-90,width=\linewidth]{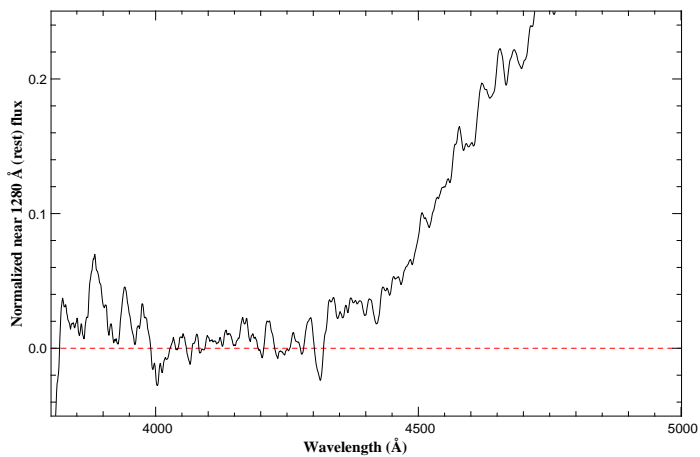}}
\caption{Result of the stacking of SDSS-DR7 spectra with a damped \lya\ system at an absorption redshift larger than 3.7 and 
a column density  $N$(H~{\sc i})~$\geq 10^{20.5} \cm ^2$. The spectra are normalized to 1 near 1280 \AA\ (in the quasar restframe). 
Due to the presence of the DLA, the flux is expected to be equal to zero at observed wavelength smaller than 4280~\AA. 
It can be seen from the figure that this is not the case in the very blue of the spectrum 
($\lambda _{\rm obs} \leq $ 4000 \AA\ ) where the mean flux is increasing. Consequently, pixels at wavelengths below 4000\AA\ are 
not used in this analysis.}
\label{fig:FluxCalibBlue}
\end{figure}

We first selected spectra with a signal-to-noise ratio per pixel greater than 
14 redwards to the \lya\ emission (the SNR is computed around 1280~\AA\ in the restframe).
We require the redshift of the quasars to be greater than 2.82 and lower than 3.00. The lower limit is chosen as such 
for two reasons: the \lya\ forest has to be complete ($z>2.7$) and we noticed that there is some 
issues with the flux calibration at the very blue end of the SDSS spectra so that we would like to avoid this part 
of the spectrum. To illustrate these problems and estimate the exact wavelength at which to start the study, we have selected spectra 
where a damped \lya\ system (DLA) is observed with aborption redshift greater than 3.7 and with a column density  
$N$(H~{\sc i})~$\geq 10^{20.5} \cm ^{-2}$ \citep[the list is available in][] {pasquieral09}.  
In these spectra, and due to the presence of the DLA, the flux is expected to be equal to zero for 
$\lambda _{\rm obs} \leq 4280$ \AA . When stacking the selected lines of sight (\Fig{FluxCalibBlue}), we note instead that 
the flux increases for wavelength lower than 4000~\AA. The difference with zero is as high as 0.05 for a normalized spectrum,
meaning that this part of the spectrum should probably not be used for the analysis. Since the study of the forest is usually
limited to beyond the O~{\sc vi} emission line ($\lambda_{\rm rest}>1050$~\AA), the minimum emission redshift will be 2.82.
The upper limit is a compromise between the number of spectra needed for the analysis and our ability to estimate confidently
a continuum by eye: it has been set to $z=3$. BAL quasars, lines of sight containing a DLA or any spectrum for which 
fitting a continuum is too risky because of missing pixels or reduction issues are removed from the analysis. 
The SDSS spectra are observed with two cameras, one for the blue and one for the red
part of the spectrum and we were concerned about the presence of a possible discontinuity or break at the 
merging point. We have avoided any spectra possibly affected by this.
When applying all those constraints, 78 spectra remain in the training set (\Tab{sample}).\\

\begin{table*}
\begin{center}
\begin{tabular}{|l|c|c|c|c||l|c|c|c|c|}
\hline
QSO & $z_{\rm em}$ & $m_{\rm g}$ & SNR & SNR & QSO & $z_{\rm em}$ & $m_{\rm g}$ & SNR & SNR \\ 
    &             &      &  (\textit{forest}) &  (\textit{red}) &     &             &      &  (\textit{forest}) &  (\textit{red}) \\
\hline
\hline
 J090439.50+221050.3 & 2.823 & 18.23 & 12.14 & 17.62 & J134826.65+290623.0 & 2.822 & 19.21 & 13.71 & 18.16 \\
 J123549.46+591027.0 & 2.820 & 17.05 & 21.16 & 28.86 & J221936.37+002434.0 & 2.850 & 18.98 & 10.73 & 17.04 \\
 J020950.71-000506.4 & 2.848 & 17.03 & 19.00 & 29.13 & J103249.88+054118.3 & 2.841 & 17.37 & 19.05 & 28.40 \\
 J121723.67+254910.5 & 2.834 & 18.50 & 10.55 & 14.09 & J143455.38+035030.9 & 2.845 & 18.35 & 9.38 & 14.94 \\
 J093434.74+543109.4 & 2.836 & 18.81 & 9.22 & 14.75 & J143823.51+083128.6 & 2.834 & 18.39 & 10.82 & 17.35 \\
 J113429.70-011153.6 & 2.860 & 19.37 & 8.10 & 14.09 & J162331.14+481842.1 & 2.830 & 17.99 & 12.94 & 19.75 \\
 J102734.13+183427.5 & 2.832 & 18.03 & 28.31 & 34.12 &  J125251.95+202405.0 & 2.877 & 18.80 & 10.47 & 15.03 \\
 J084948.98+275638.7 & 2.845 & 17.96 & 14.99 & 21.35 & J131114.90-023045.5 & 2.860 & 18.51 & 13.07 & 16.97 \\
 J123503.01+441118.2 & 2.845 & 18.54 & 8.18 & 14.25 & J124652.81-081312.9 & 2.854 & 18.27 & 18.04 & 30.13 \\
 J094152.90+621705.5 & 2.862 & 19.48 & 11.07 & 15.44 & J012156.03+144823.9 & 2.869 & 17.22 & 25.65 & 31.14 \\
 J134450.86+191843.4 & 2.860 & 19.68 & 11.20 & 14.92 & J102832.09-004607.0 & 2.867 & 18.14 & 9.91 & 15.25 \\
 J160843.90+071508.6 & 2.877 & 16.80 & 25.54 & 35.30 & J103000.37+440010.7 & 2.870 & 18.99 & 11.72 & 16.81 \\
 J031522.09-080043.7 & 2.895 & 18.42 & 11.68 & 18.74 & J114218.85+572824.4 & 2.897 & 18.82 & 9.86 & 15.14 \\
 J105236.34+253956.1 & 2.896 & 17.85 & 18.08 & 23.63 & J093207.46+365745.5 & 2.878 & 17.70 & 22.55 & 27.03 \\
 J111033.39-113259.9 & 2.895 & 18.17 & 11.38 & 17.02 & J112857.85+362250.2 & 2.885 & 18.04 & 15.51 & 20.37 \\
 J085733.39+370045.6 & 2.946 & 18.94 & 15.53 & 22.39 & J135044.66+571642.9 & 2.893 & 17.45 & 19.79 & 27.28 \\
 J170144.31+284234.2 & 2.913 & 19.28 & 11.54 & 18.79 &  J003311.34-171041.5 & 2.905 & 17.97 & 12.08 & 20.60 \\
 J124754.78+492758.2 & 2.935 & 19.13 & 15.06 & 22.02 & J131440.18+271449.5 & 2.935 & 18.68 & 10.86 & 14.69 \\
 J101539.35+111815.9 & 2.905 & 18.57 & 12.10 & 16.10 & J094442.31+255443.3 & 2.901 & 18.59 & 11.60 & 16.48 \\
 J142950.91+260750.2 & 2.910 & 18.40 & 11.57 & 16.77 & J093643.51+292713.6 & 2.922 & 18.10 & 13.79 & 17.98 \\
 J163842.52+360213.8 & 2.910 & 19.36 & 13.18 & 19.67 & J120322.71+403310.1 & 2.915 & 19.14 & 11.10 & 15.88 \\
 J012305.58+063047.2 & 2.923 & 19.07 & 11.99 & 16.07 & J135225.88+293830.4 & 2.915 & 18.43 & 9.32 & 14.35 \\
 J102807.74+172956.8 & 2.926 & 18.11 & 30.59 & 41.45 & J100808.27+285214.6 & 2.921 & 18.53 & 10.11 & 14.56 \\
J130554.81+184904.9 & 2.925 & 18.27 & 13.79 & 21.28 & J152119.68-004818.7 & 2.934 & 17.93 & 19.39 & 25.50 \\
J075326.11+403038.6 & 2.929 & 17.92 & 13.80 & 20.90  & J103928.63+253345.4 & 2.933 & 18.69 & 11.40 & 14.90 \\
J004129.80+241702.1 & 2.934 & 19.32 & 12.94 & 18.51 & J223927.69+230018.0 & 2.928 & 18.41 & 14.47 & 19.71 \\
J160441.47+164538.3 & 2.932 & 16.91 & 29.71 & 38.00 & J102025.27+334633.4 & 2.930 & 18.21 & 13.54 & 18.48 \\
J120753.79+325747.4 & 2.945 & 18.89 & 9.92 & 14.25  & J141442.96+193523.5 & 2.946 & 18.32 & 11.46 & 15.61 \\
J131212.60+001129.7 & 2.945 & 19.27 & 9.76 & 16.18 & J082257.04+070104.3 & 2.940 & 18.65 & 11.05 & 15.02 \\
J111038.63+483115.6 & 2.953 & 16.79 & 26.22 & 33.02 & J130337.21+194926.7 & 2.953 & 17.84 & 16.79 & 22.49 \\
J104253.44-001300.8 & 2.953 & 18.87 & 11.98 & 17.11 & J134811.76+281801.8 & 2.966 & 17.57 & 26.29 & 39.35 \\
J091546.67+054942.7 & 2.967 & 18.40 & 9.35 & 14.37 & J125708.23+191857.2 & 2.970 & 18.61 & 19.54 & 27.02 \\
J113559.41+422004.4 & 2.953 & 18.57 & 9.35 & 14.38 & J090423.37+130920.7 & 2.968 & 17.59 & 22.46 & 30.62 \\
J120331.29+152254.7 & 2.976 & 16.99 & 24.55 & 32.71 & J142807.87+162634.3 & 2.977 & 18.32 & 12.90 & 18.00 \\
J013829.75+224558.9 & 2.987 & 19.11 & 10.03 & 14.84 & J085959.14+020519.7 & 2.970 & 18.45 & 10.52 & 15.43 \\
J132255.66+391207.9 & 2.984 & 17.76 & 19.45 & 25.44 & J120006.25+312630.8 & 2.978 & 16.62 & 29.81 & 34.04 \\
J132321.24+250027.4 & 2.980 & 18.00 & 11.72 & 15.97 & J125419.07+362750.4 & 2.980 & 18.51 & 13.68 & 17.43 \\
J074313.86+28442.3 & 2.975 & 19.45 & 12.31 & 14.84 & J143912.34+295448.0 & 2.992 & 17.65 & 17.65 & 24.10 \\
J075710.36+362301.5 & 2.990 & 18.79 & 11.63 & 15.36 & J224154.38+000102.1 & 2.998 & 19.12 & 9.57 & 14.26 \\
\hline
\end{tabular}
\end{center}
\caption{List of SDSS-DR7 quasars used to define the correlation matrix and the eigenvectors at $z\sim 3$}
\label{t:sample}
\end{table*}%

\begin{figure*}
\centering{\includegraphics[angle=-90,width=130mm]{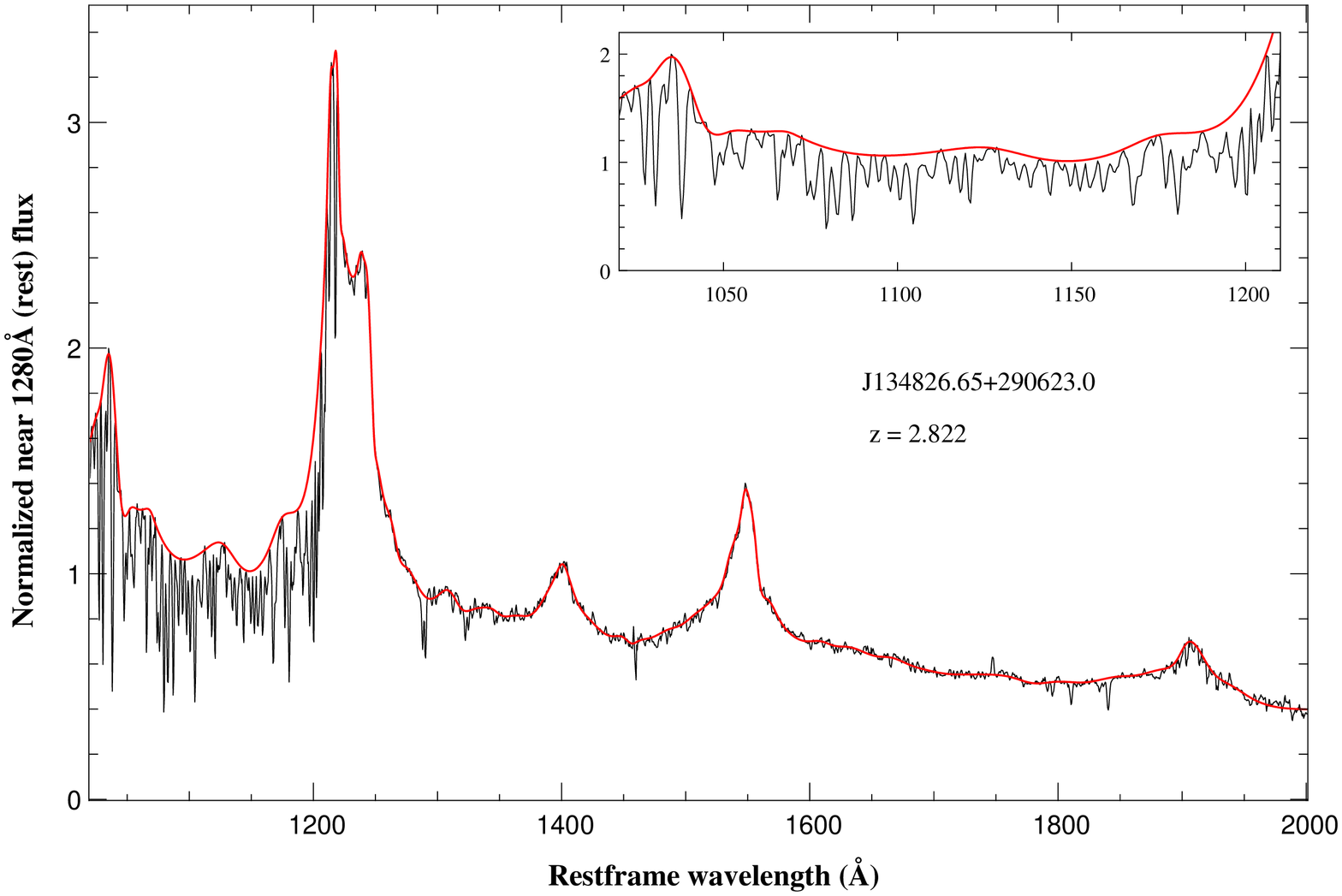}}
\caption{Spectrum and continuum of the quasar SDSS J134826.65+290623.0. This quasar belongs to the sample of SDSS $z\sim3$ quasars 
that is used to derive the Principal Component Analysis eigenvectors (\Sec{pcaz3}). Our estimate of the continuum is shown with the thick red line 
and a zoom in the Lyman-$\alpha$ forest region is shown in the inset. }
\label{fig:DR7spectra}
\end{figure*}

\begin{figure}
	\centering{\includegraphics[width=\linewidth]{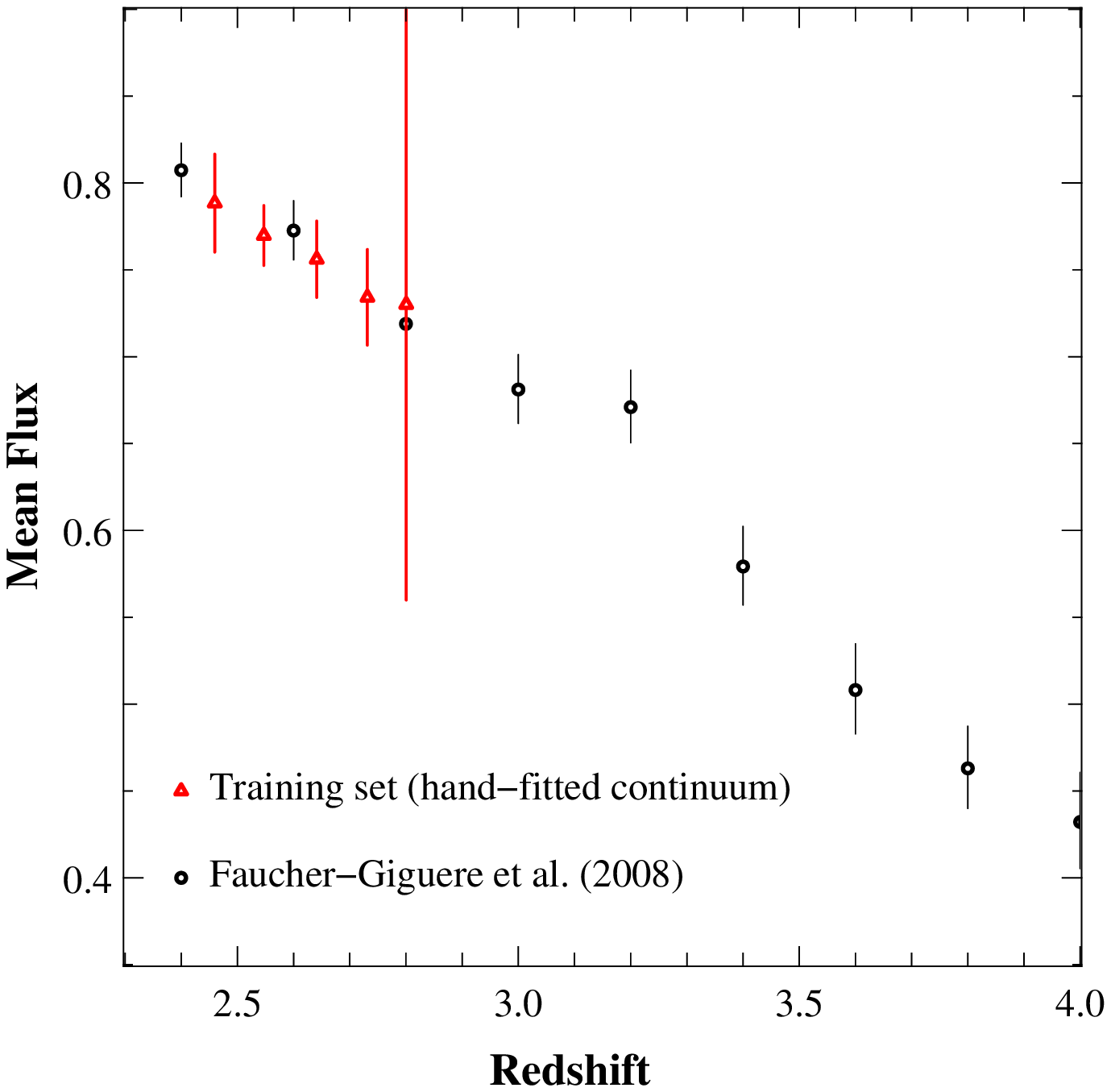}}
\caption{The mean flux evolution in the training set (with redshift bins of size $\Delta z=0.1$; red triangles) is compared 
to FG08 measurements from high resolution and high signal-to-noise spectra (black circles). The continuum of the 78 SDSS spectra 
in the training set have been fitted by hand using spline interpolation. Error bars from our measurements have been computed by bootstrapping pixels 
of our sample. Both measurements are consistent, making us confident in our continuum estimate.}
\label{fig:MeanFluxTS}
\end{figure}

Once the training set is defined, spectra are smoothed  and the continuum is "hand-fitted". Redwards to the \lya\ emission line, we 
follow the different emission lines and ignore isolated absorption lines. In the \lya\ forest, the continuum cannot be 
uniquely defined. We assume that the continuum has a smooth shape, that it roughly follows the peaks of the spectrum, and that the blending 
of lines at the SDSS resolution is large. Points are placed around the peaks of the flux, and a spline interpolation is used to connect 
them. After a first try, we minimize the number of  points used and we check that the continuum is indeed located above blends 
of lines, but at about the level of 'flat' regions. A typical example is given in \Fig{DR7spectra}. 
Note that with this procedure, we {\sl de facto} take into account that the chosen points can be affected by some absorption.
To check that this is indeed the case, the spectra are then rebinned with 0.5 \AA\ restframe pixels and the mean flux 
evolution is computed and compared to \cite{FG2008} measurements from
high and medium resolution and high signal-to-noise spectra. Both estimates are in agreement giving us confidence in our 
hand-fitted continua (\Fig{MeanFluxTS}).\\

\begin{figure*}
	\centering{\includegraphics[width=130mm]{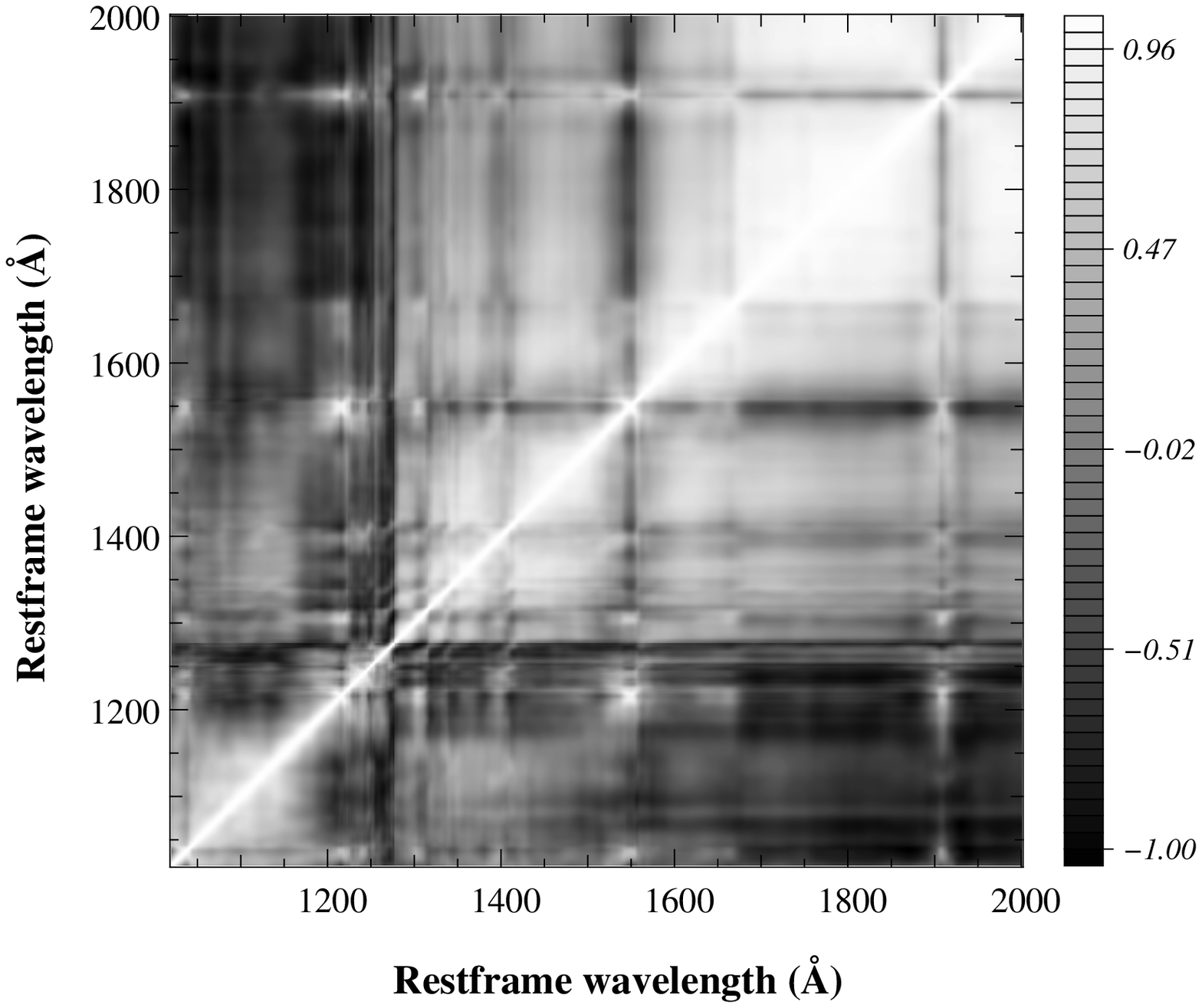}}
\caption{Correlation matrix computed with the training set. A moderate correlation (0.3-0.6) is found between the shape of the continuum 
in the \lya\ forest (1020 $\leq \lambda \leq$1210 \AA) and in the region between \lya\ and \CIV\ emission lines 
(1216 $\leq \lambda \leq$1600 \AA), in agreement with S05. We note also a moderate anti-correlation (from -0.6 to -0.4) 
between the continuum in the \lya\ forest and the region further to \CIV\ emission line.} 
\label{fig:CorrelationMatrix}
\end{figure*}

The procedure described in \Sec{pca} is then performed and the correlation matrix is computed and displayed in \Fig{CorrelationMatrix}. 
In agreement with S05, a moderate correlation (0.3-0.6) is found between the shape of the continuum in the forest and the region between 
\lya\ and \CIV\ emission lines. Thanks to the larger restframe wavelength coverage of this study, a moderate anti-correlation 
(from -0.6 to -0.4) between the shape of the continuum in the forest and in the region between \CIV\ and \CIII\ emission lines is found. 
S05 noticed that a PCA continuum has the good shape in the forest but that the amplitude of the power-law component is unstable. The 
anti-correlation in that extra-part of the continuum may improve the stability of the prediction of the amplitude of the continuum in 
the forest. This is discussed in more details in \Sec{bootstrapTrainingSet}.\\

\label{s:results} 

\begin{figure*}
\centering{\includegraphics[angle=-90,width=130mm]{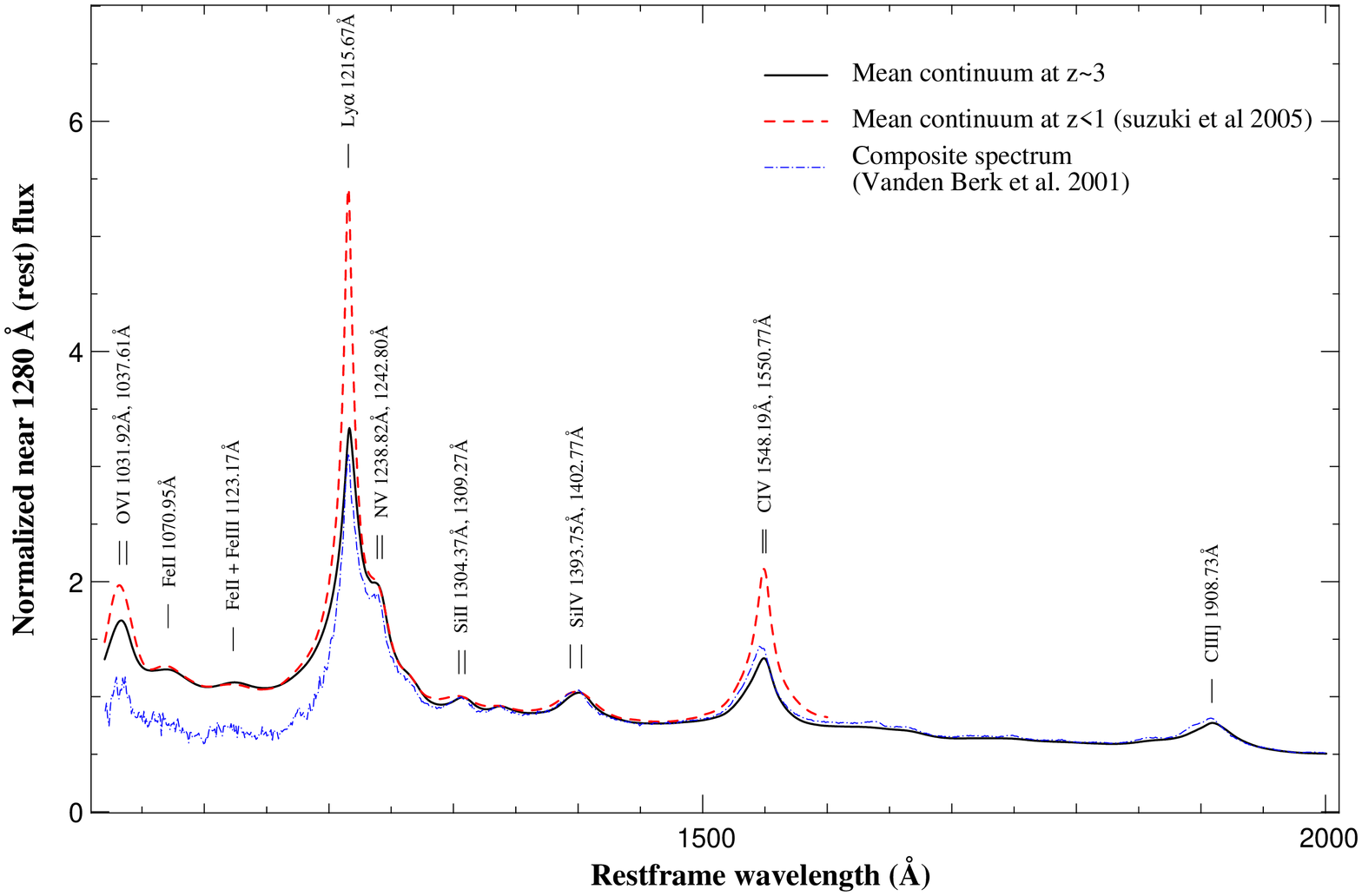}}
\caption{The solid black (resp. dashed red) line is the mean continuum of quasars at $z=3$ (resp. $z \leq 1$, S05). The wavelength coverage at  $z=3$ corresponds to SDSS spectra and is larger than at $z \leq 1$. The main difference between the two
mean spectra is seen in the amplitude of emission lines. 
The composite spectrum from \cite{VB01} computed from 2,200 SDSS spectra (dash-dot blue line) is in good agreement 
with our mean continuum. The difference in the Lyman-$\alpha$ forest is due to the fact that \cite{VB01} did not try to avoid
the absorption from the IGM.  
A small shift in the position of emission lines can be noticed: this is because \cite{VB01} have used the Mg~{\sc ii}
line as a reference to compute the redshift whereas we have used the \CIV\ and \CIII\ lines.}
\label{fig:PCAcont}
\end{figure*}

The mean continuum is shown in \Fig{PCAcont} together with the mean continuum of S05 ($z \leq 1$) and the composite spectrum 
derived by \cite{VB01}.
In the wavelength range of interest here, quasars contributing to the \cite{VB01} composite cover a redshift range from 2.13 to 4.789 
for the \lya\ region and from 1.5 to 4.789 for \CIV. Our mean continuum is in excellent agreement redwards to the \lya\ emission line
with the \cite{VB01} composite. The discrepancy in the blue is 
simply due to the absorption in the \lya\ forest that \cite{VB01} did not try to remove.
When comparing to S05, one can see that the amplitude of \CIV, \lya\ and \lyb\ emission lines relative
to the continuum are less important in the SDSS spectra. While the determination of the \CIV\ emission is relatively straightforward, 
the presence of absorption at the position of the \lyb\ emission line and in the blue wing of the \lya\ emission line makes
the continuum difficult to estimate. Thus, the main and robust difference between the mean continua in SDSS and HST spectra is 
the variation of the \CIV\ equivalent width.
Such an evolution of the QSO emission line equivalent widths has been noted for long  \citep{baldwin1977} and
has also been reported by \cite{zheng97}. The later authors used 101 HST spectra to compute a low-$z$ composite spectrum (90\% of the quasars 
had a redshift lower than 1.5) and compared it to the \cite{francis91} composite ($z \sim 3$). This evolution is probably related to 
the quasar luminosity. Note that no evolution is found by \cite{fan2009} from $z \sim 2$ to $z>6$.

\begin{figure*}
	\centering{\includegraphics[width=\linewidth]{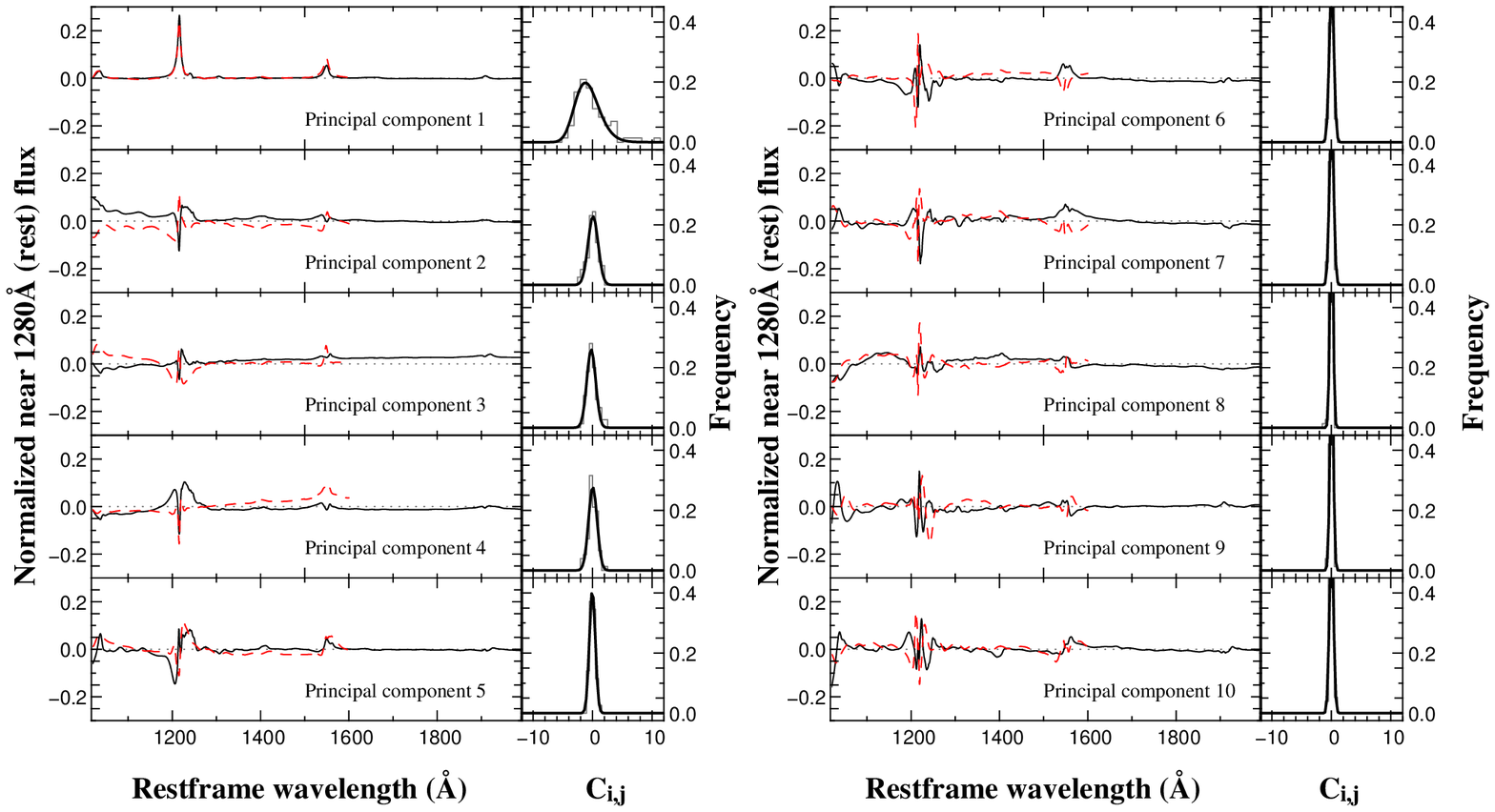}}
\caption{First ten principal components of a Principal Component Analysis applied to (i) $z\sim 3$ 
SDSS quasar spectra (black solid lines) over the range 1020$-$2000 \AA\ and to (ii)  $z \leq 1$ HST quasar spectra (S05, red dashed lines) over the range 1020$-$1600 \AA\ . 
The distributions of the coefficient associated to each component are shown in the right panel (grey histogram) together with their 
fit with a Gaussian (except for the first component for which the distribution is log-normal; thick black line).}
\label{fig:PCAcomponent}
\end{figure*}

\begin{figure*}
	\centering{\includegraphics[width=\linewidth]{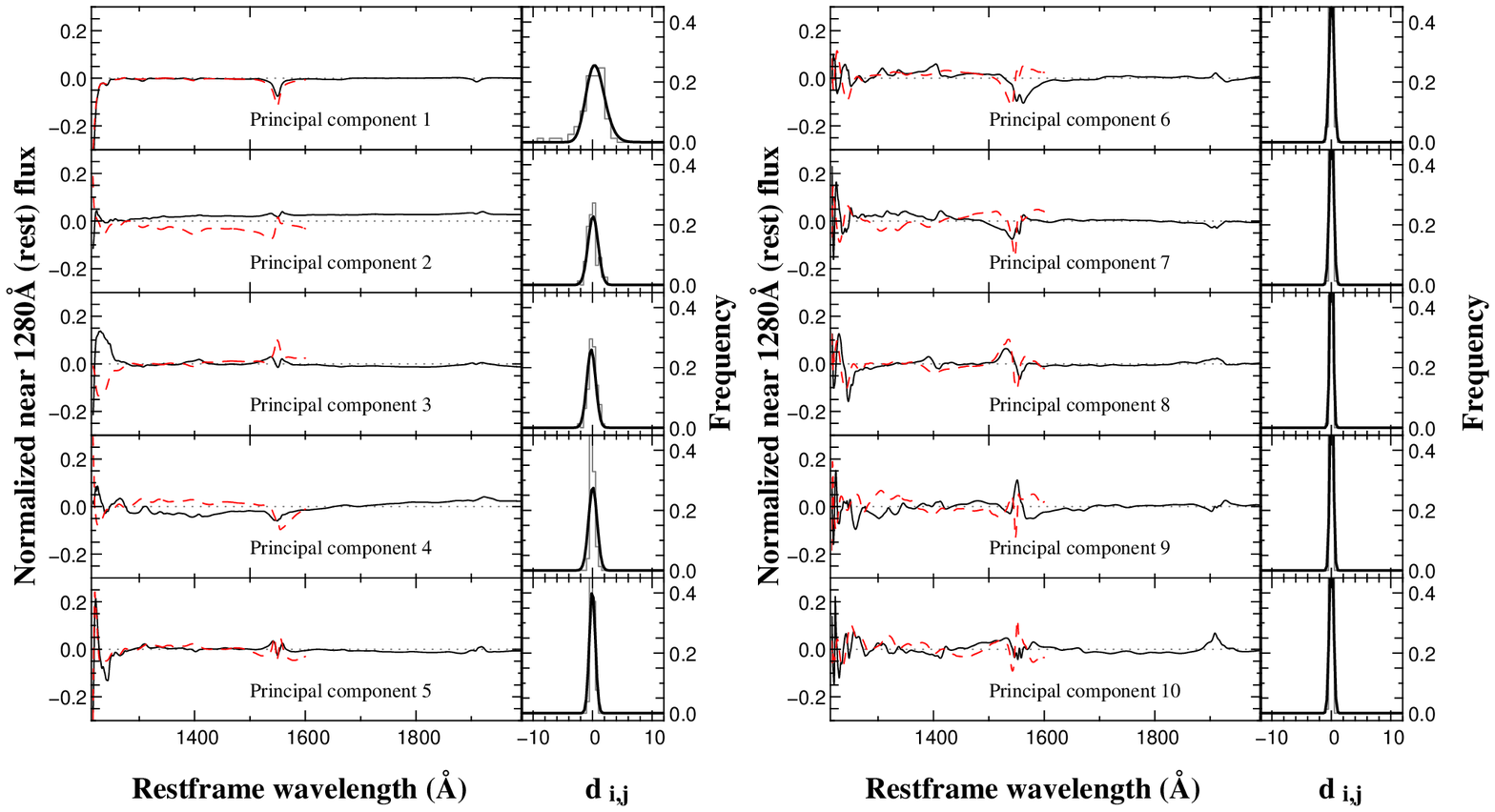}}
\caption{First ten principal components obtained from a Principal Component Analysis of the QSO spectrum redwards to the \lya\ emission line, 
for (i) the $z\sim 3$ SDSS quasar spectra (black solid lines) and (ii) the $z \leq 1$ HST quasar spectra (S05, red dashed lines). The distributions 
of the coefficient associated to each component are shown in the right panels (grey histogram) together with a Gaussian fit 
(except for the first component for which the distribution is log-normal; thick black line).}
\label{fig:PCAcomponentRed}
\end{figure*}

Further, the first ten eigenvectors are displayed in \Fig{PCAcomponent} when derived from the full wavelength coverage 
and in \Fig{PCAcomponentRed} when derived from the region redwards to the \lya\ emission, together with the components
provided by S05. The distributions of the coefficients, $c_{\rm i,j}$ and $d_{\rm i,j}$, computed on our sample using 
\Eq{eq_weight}, are also shown.
The first component looks very similar in the two decompositions and is dominated by the amplitude of the \lya\ and 
\CIV\ emission lines. The most important difference with S05 lies in the shape of the distribution of the associated coefficients: 
in their study, the distribution is Gaussian whereas in ours, this distribution is log-normal. This trend (concerning the 
eigenvectors and the distributions of coefficients) is in agreement with what has been found by \cite{francis92}.
The discrepancy between the shapes of the coefficient distributions could mean that the S05 sample is more homogeneous 
than ours in term of amplitude of emission lines.\\
The second component is dominated by the continuum slope and it seems that there is a difference between what is found here 
and in S05 which seems to be somewhat compensated by the difference in the third component. 
Other components show small differences but they are less pronounced and the coefficient distributions are very similar.\\
To estimate more quantitatively the similarity of the low and high redshift sets of eigenspectra, 
we follow \cite{Yipal04}, and compute the sum of the projection operators of each set of eigenvectors $|\xi_{\rm j}>$:
\begin{eqnarray}
 \mathbf{\Xi} = \sum_{\rm j=1,m} |\xi _{\rm j}> <\xi_{\rm j}| \ ,  
\end{eqnarray}
\noindent
and then the trace of the products of the projection operators:
\begin{eqnarray}
 Tr(\mathbf{\Xi _{z=3} \Xi _{S05} \Xi _{z=3}}) & = & D \ ,
\end{eqnarray} 
where D will be the common dimension of both sets.  
\noindent 
The two sets are disjoint if the trace is zero. If the basis are completely alike, $D$ should be equal to their dimension, therefore
$D=10$ in our case. 
To compute this number, we cut our eigenspectra at 1600 \AA\ (rest) and we find $D = 7.6$.
This means that the two decompositions are similar but not exactly the same, confirming the slight evolution of the decomposition
with redshift.

We provide in an electronic form the first 10 eigenvectors of the PCA. The distributions of associated coefficients are very close to gaussian functions, 
except for the first coefficient, the distribution of which is fitted with a log-normal distribution. 
Their characteristics are listed in \Tab{coef_fit}.

\begin{table}
\begin{center}
\begin{tabular}{|c|c|c|c|c|}
\hline
Component $j$ & \multicolumn{4}{c|}{$c_{i,j}$ distribution} \\ 
  & $ \mu $ & $ \sigma _{\mu } $ & $ \sigma $  & $ \sigma _{\sigma }$ \\
\hline
\hline
1 & 2.645 & 0.008 & 0.145 & 0.006 \\
2 & 0.090 & 0.033 & 0.845 & 0.033 \\
3 & -0.220 & 0.019 & 0.745 & 0.019 \\
4 & 0.044 & 0.031 & 0.717 & 0.031 \\
5 & -0.029 & 0.003 & 0.496 & 0.004 \\
6 & 0.028 & 0.002 & 0.386 & 0.003 \\
7 & 0.020 & 0.003 & 0.363 & 0.004 \\
8 & 0.018 & 0.002 & 0.258 & 0.008 \\
9 & 0.034 & 0.002 & 0.284 & 0.006 \\
10  & 0.008 & 0.001 & 0.336 & 0.001 \\
\hline
\end{tabular}
\end{center}
\caption{Parameters of the fits to the distributions of weights, $c_{\rm i,j}$, for the first ten principal components. The distributions 
have been fitted with gaussian functions, except for the distribution of the first coefficient which has been fitted with a 
lognormal function.}
\label{t:coef_fit}
\end{table}%

\subsection{Quality of the predicted continuum}
\label{s:validation}

The decomposition of the quasar emission investigated at $z \leq 1$ by S05 with HST spectra and at $z \sim 3$ in this paper with 
SDSS spectra yield two similar but different basis of eigenvectors, as shown in the previous Section. One would like 
to know if the larger wavelength coverage of our eigenvectors provides any advantage and how far the 
new determination is required to reproduce the correct quasar continuum at $z \sim 3$. In other words, is the prediction of 
quasar continuum at $z \sim 3$ better if one uses the new PCA eigenvectors derived in this paper ?
To answer this question, we apply three tests to the predicted continua.

\subsubsection{Error on the predicted PCA continuum in the Lyman-$\alpha$ forest}
\label{s:bootstrapTrainingSet}

In order to estimate the difference between the true and predicted continuum in the Lyman-$\alpha$ forest, 
a set of eigenvectors and a projection matrix are derived using 77 spectra of the training set (out of 78)
and the continuum over the Lyman-$\alpha$ forest for the remaining quasar is estimated using these 
parameters and following the method described in \Sec{pca}.
That procedure is repeated on each of the 78 spectra in the training set.\\
Following S05, we estimate for each spectrum, the absolute fractional flux error 
$\left | \delta F \right |$, defined as follows,
\begin{equation}
\left | \delta F \right | =
\left .
\int_{\lambda_1}^{\lambda_2}
\left | \frac{p(\lambda)-q(\lambda)}{q(\lambda)} \right | 
d \lambda 
\right /
\int_{\lambda_1}^{\lambda_2} d \lambda ,
\label{eq:eq_fluxerror}
\end{equation}
where $p$ and $q$ are, respectively, the predicted and real continua.

\begin{figure}
	\centering{\includegraphics[width=\linewidth]{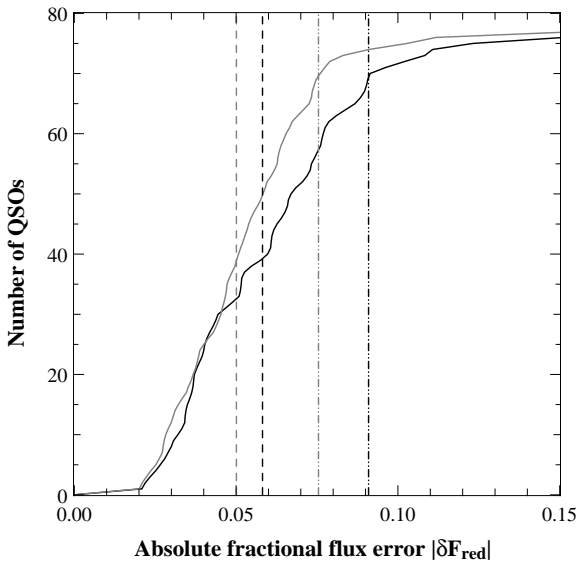}}
\caption{Cumulative distributions of the absolute fractional flux error (\Eq{eq_fluxerror}) redwards to the \lya\ emission line 
when predicting spectra with PCA decompositions over a wavelength range extending up to
2000~\AA\ (black solid line) or up to 1600~\AA\ (grey solid line). Median values are similar for the two estimates 
(black and grey dashed vertical lines respectively). The 90$^{th}$ percentile from the 1600~\AA\ decomposition (dash-dot-dot grey vertical line) is 
lower than one for the 2000~\AA\ decomposition (dash-dot-dot black vertical line).}
\label{fig:AbsErrorRed}
\end{figure}

\begin{figure}
	\centering{\includegraphics[width=\linewidth]{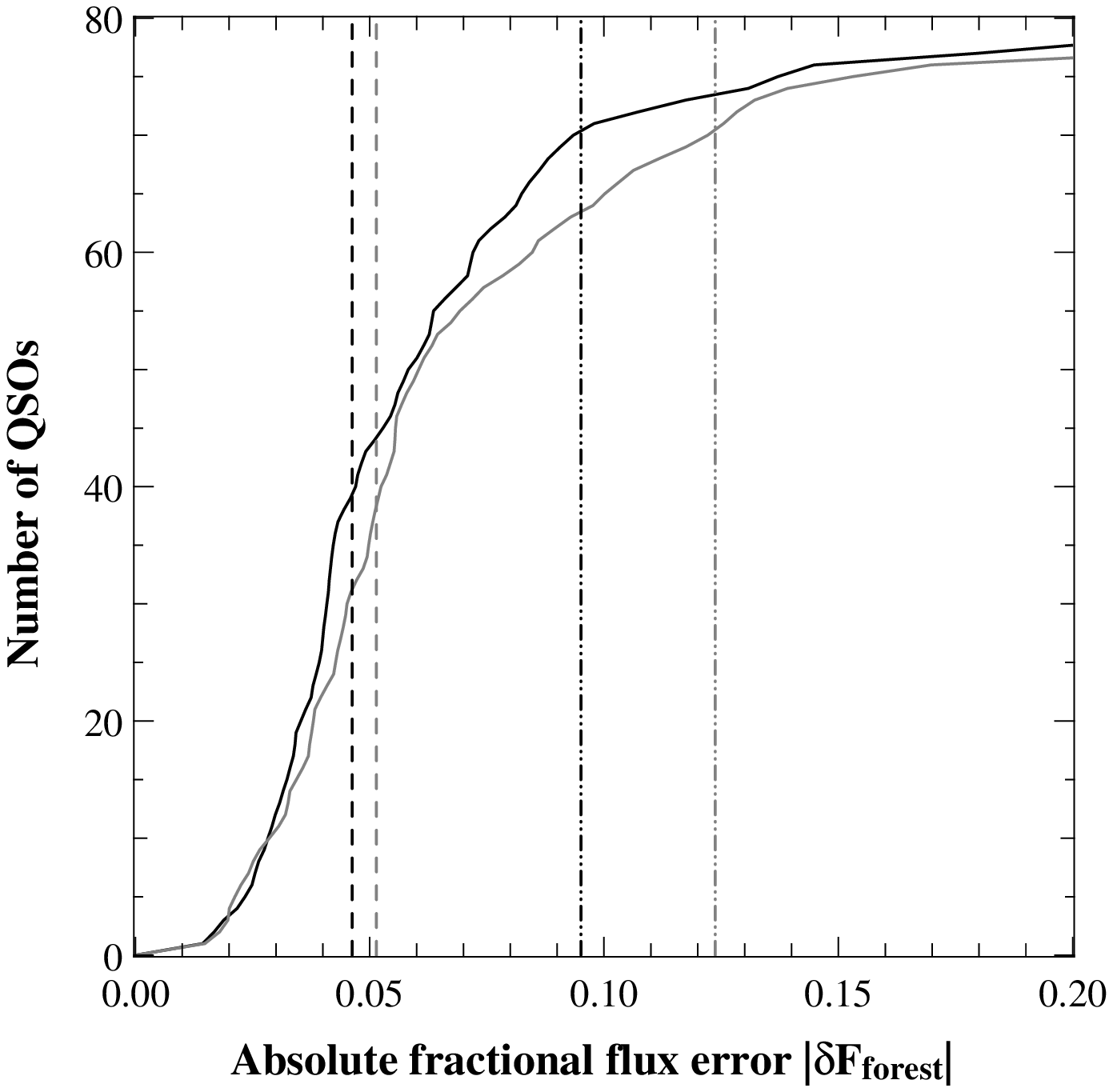}}
\caption{Same as \Fig{AbsErrorRed} for the cumulative distributions of the absolute fractional flux error 
(\Eq{eq_fluxerror}) in the \lya\  forest. }
\label{fig:AbsErrorBlue}
\end{figure}

This has been computed over the restframe wavelength ranges, 1050-1170~\AA\ in the forest, and 1280-2000~\AA\ (or 1280-1600~\AA) in the red. 
The cumulative distributions of the absolute fractional flux errors are plotted in \Fig{AbsErrorRed} 
(in the red) and \Fig{AbsErrorBlue} (in the forest) using two different wavelength coverages (black line for 2000~\AA\ and grey line for 1600~\AA).\\
In the red, the median error is 5.8\% and 5\% using, respectively, the 2000~\AA\ and the 1600~\AA\ decompositions (\Fig{AbsErrorRed}, black dashed and 
grey dotted vertical lines respectively) and the 90$^{th}$ percentile error with the 2000~\AA\ decomposition is larger than the error using 
the 1600~\AA\ set of eigenspectra. This indicates simply that the range up to 1600~\AA~ is easier to fit.

\begin{figure}
\vbox{
	\centering{\includegraphics[angle=-90,width=\linewidth]{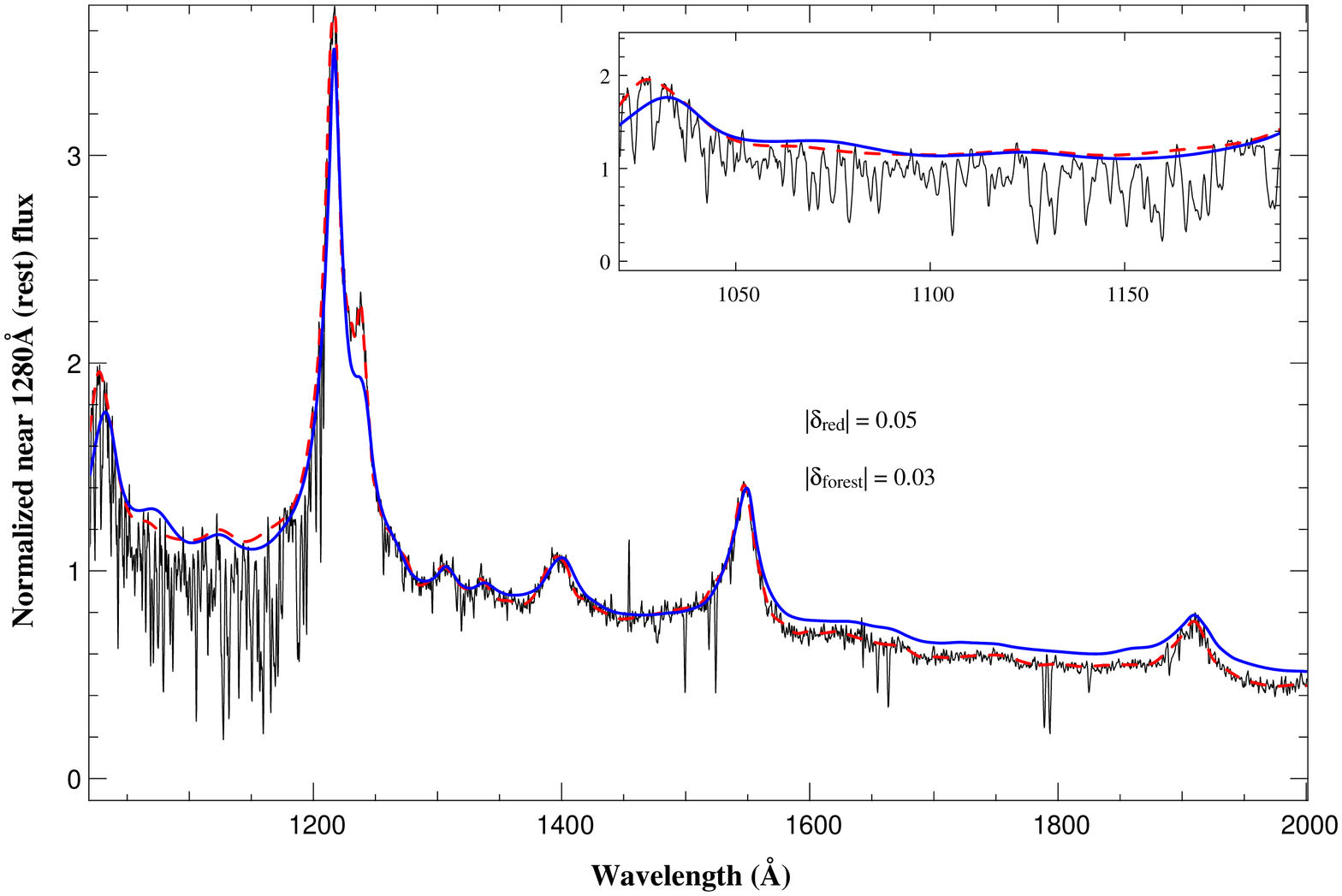}}
        \centering{\includegraphics[angle=-90,width=\linewidth]{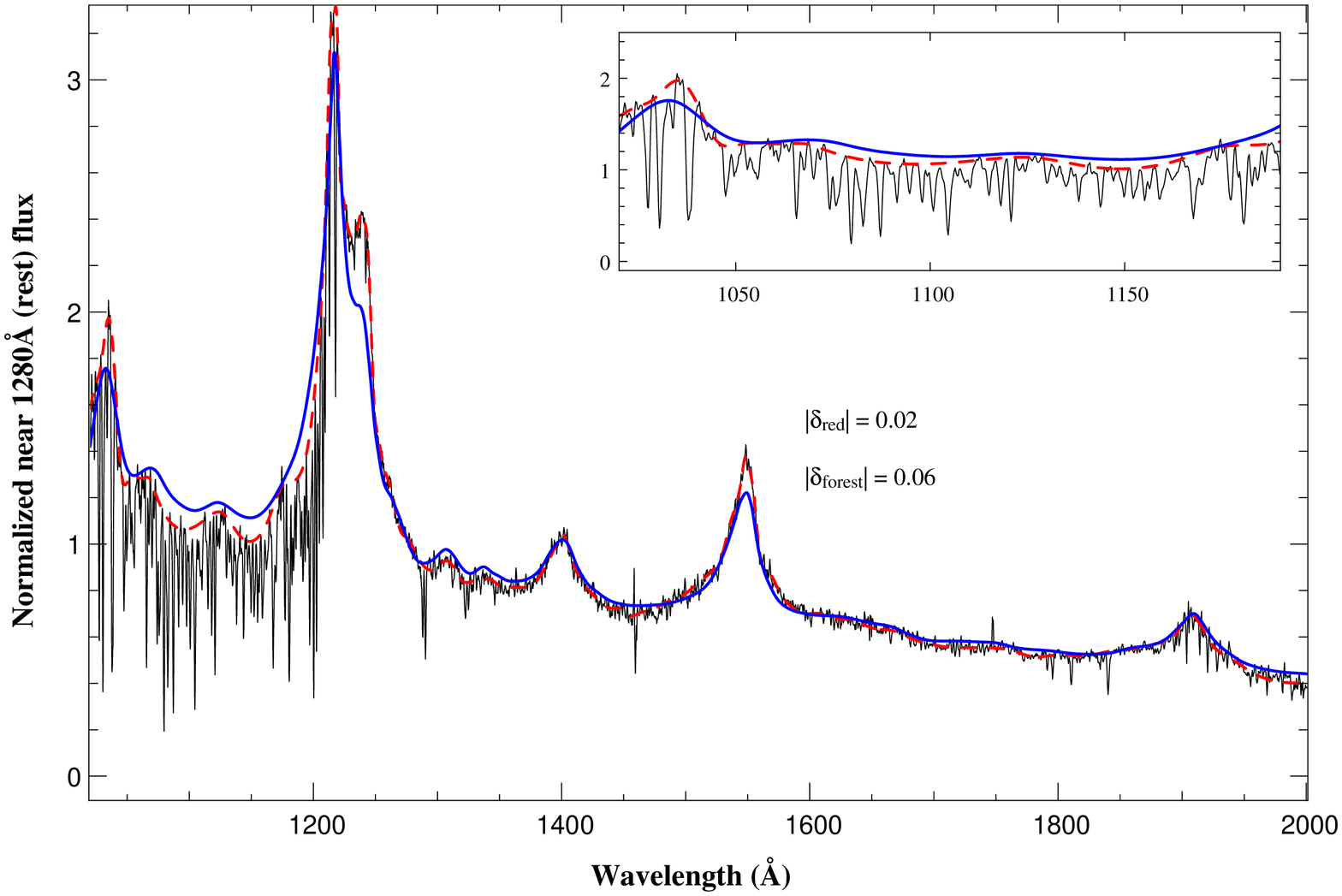}}
}
\caption{Example of a spectrum with (i) a large absolute fractional flux error in the red and a small error in the forest
(upper panel) and (ii) a large absolute fractional flux error in the forest and a small error in the red (lower panel). 
The hand-fitted continuum is the red dashed line and the predicted one is the solid blue line.}
\label{fig:ExErr1}
\end{figure}

The important result is that the opposite trend is observed in the forest (\Fig{AbsErrorBlue}): median values (around 5\%) are very close 
for the two wavelength coverages (black and grey  dashed vertical lines) whereas the 90$^{th}$ percentiles are different (black and grey dot-dot-dash vertical lines) 
with 9.5\% and 12.4\% errors for, respectively, the 2000 and 1600~\AA\ decompositions. This means that using the full coverage reduces the number of outliers with more 
than 12\% error in the \lya\ forest. Errors in S05 are similar to those we find here but, by using the 2000\AA\ decomposition, outliers are less frequent than 
in the previous study. To illustrate what those numbers mean on the continuum level, examples of spectra with their predicted continua 
are displayed in \Fig{ExErr1}.

\subsubsection{Distribution of spectral indices}
\label{s:mock-quality}
\begin{figure}
	\includegraphics[width=.9\linewidth]{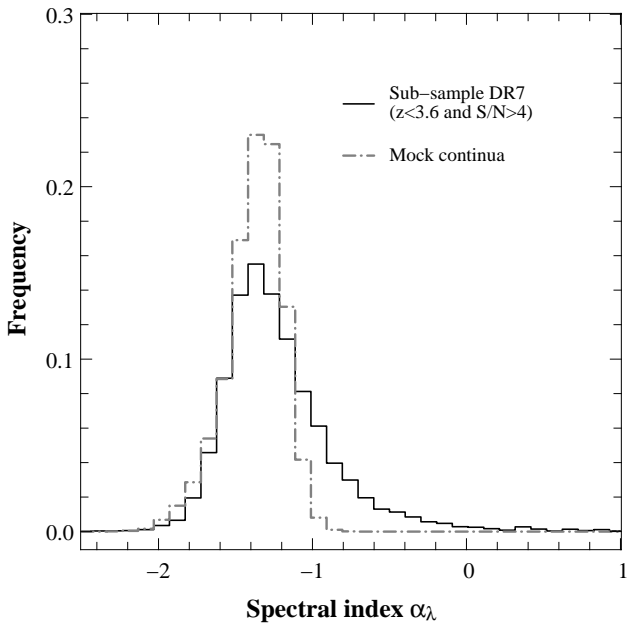}
\caption{Distribution of spectral indices derived by fitting a power-law to
mock continua generated using the principal components and the distributions of coefficients $c_{\rm ij}$ derived from 
our SDSS-DR7 subsample (grey histogram) compared to the spectral index distribution obtained from fitting SDSS-DR7 spectra (black histogram).  
}
\label{fig:alphaLambda_distri}
\end{figure}

We can also compare the characteristics of mock continua generated from the set of computed eigenvectors and the distribution 
of weights $c_{\rm ij}$ given in \Tab{coef_fit} with those of real spectra from SDSS-DR7.
For this, we have fitted in the same way a power-law to mock continua and to SDSS-DR7 spectra 
(see \Sec{DR7mf} for more details). 
The normalized (sum equal 1) distributions of the derived power-law index are displayed in \Fig{alphaLambda_distri}. The 
distribution from mock spectra is more peaked than the one from SDSS spectra but is centered around the same value. This behavior is 
expected because the PCA gives us a mean description of the whole quasar population.\\

\subsubsection{Prediction of the continuum: fitting coefficients versus using the projection matrix}
\label{s:FitOrProjection}

\begin{figure}
  \centering\includegraphics[width = \linewidth]{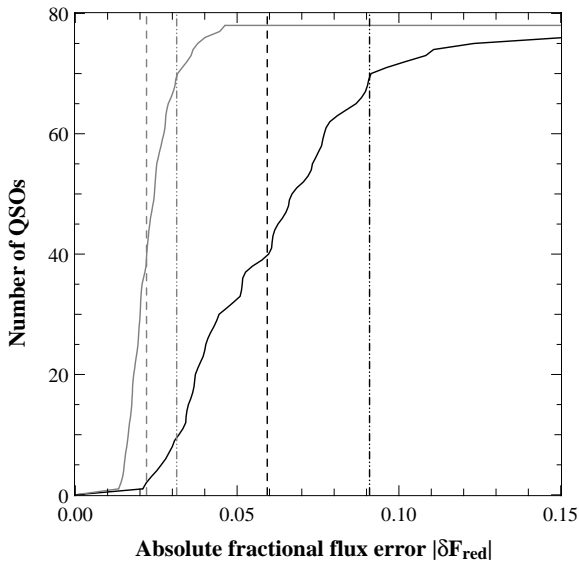}
\caption{Cumulative distributions of the absolute fractional flux error in the red part of the spectra in the training set:
(i) when the weights used to reconstruct the spectrum are those obtained from the red part of the
spectrum (method 1, grey line); 
(ii) when the projection matrix is used (method 2, black lines). Median values are displayed (dashed vertical lines) together 
with 90$^{th}$ percentiles (dot-dot-dashed vertical lines). }
\label{fig:ErrorFitProjRed}
\end{figure}

\begin{figure}
  \centering\includegraphics[width = \linewidth]{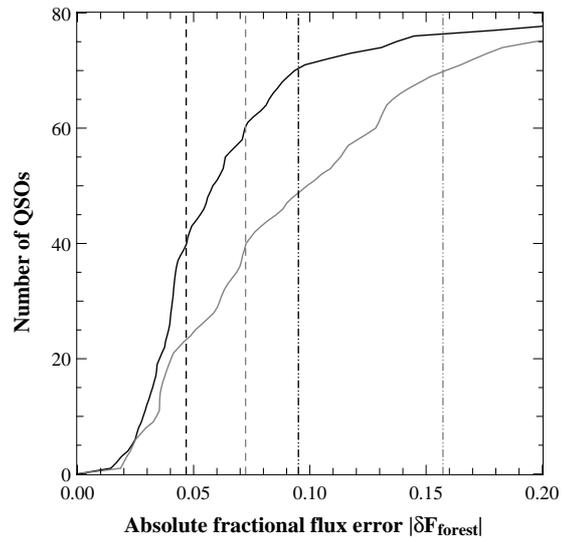}
\caption{Same as \Fig{ErrorFitProjRed} in the \lya\ forest. Errors are less when the projection
matrix is used (method 2, black line, see Text). The number of outliers (spectra with large errors) is much smaller
in that case. }
 \label{fig:ErrorFitProjBlue}
\end{figure}

Our goal is to estimate the quasar continuum over the Lyman-$\alpha$ forest. For this, we estimate first the
weights of the red part of the spectrum in the basis obtained from PCA of the red part of the spectra.
We then multiply these weights by the projection matrix (see Eq.~5) to compute the weights to be used in the basis 
obtained from the overwhole spectrum. This gives us the spectrum reconstructed in the whole
wavelength range (method 1).
One may be tempted to directly use the coefficients obtained from the red part of the spectrum as 
representative of the whole spectrum just replacing the eigenvectors 
obtained from the red part by those obtained from the full wavelength coverage (method 2). To test how useful the projection 
is, both methods have been used to predict the continuum and 
the distributions of absolute fractional flux error (\Eq{eq_fluxerror}) have been computed  and 
are displayed in \Fig{ErrorFitProjRed} and \Fig{ErrorFitProjBlue}.

When using method 2, and not surprinsigly, the red part of the spectrum is very well fitted with a median error 
less than 2.5\% (\Fig{ErrorFitProjRed}, dashed grey line) and method 1 (projection)  
leads to larger errors (dashed black line in \Fig{ErrorFitProjRed}, median error $\sim$6\%). 
In the forest, the trend is opposite with a median error less than 5\% when 
method 1 is applied (\Fig{ErrorFitProjBlue}, dashed black line) and more than 7\% when method 2 is applied (dashed grey line). 
In addition, with method 2, 10\% of the spectra have more than 15\% error (dot-dot-dash grey line) for only one percent 
with method 1. It is therefore apparent that the projection matrix should be used.

%
\section{Evolution  of the mean flux}
\label{s:meanflux} 

An important application of the quasar continuum estimate over
the Lyman-$\alpha$ forest wavelength range is the determination of the redshift evolution of the mean
flux in the IGM. Numerous authors have performed this measurement
using high and/or intermediate spectral resolution data \citep[\textit{e.g.}][]{songaila2004,bernardial2003,Dallaglioal08,Dallaglioal09,FG2008}.
The evolution is smooth except for a possible bump at $z\sim 3.2$ which could be related to the He~{\sc ii}
reionization (Schaye et al. 2000),  although this is not the only possible explanation 
\citep{FG2008}. 
In this Section we reinvestigate this issue applying the method developped in \Sec{pcaz3}.

\subsection{Comparison with B03}
\label{s:compToB03}

\begin{figure*}
\begin{minipage}[h]{.5\linewidth}
	\includegraphics[width=\linewidth]{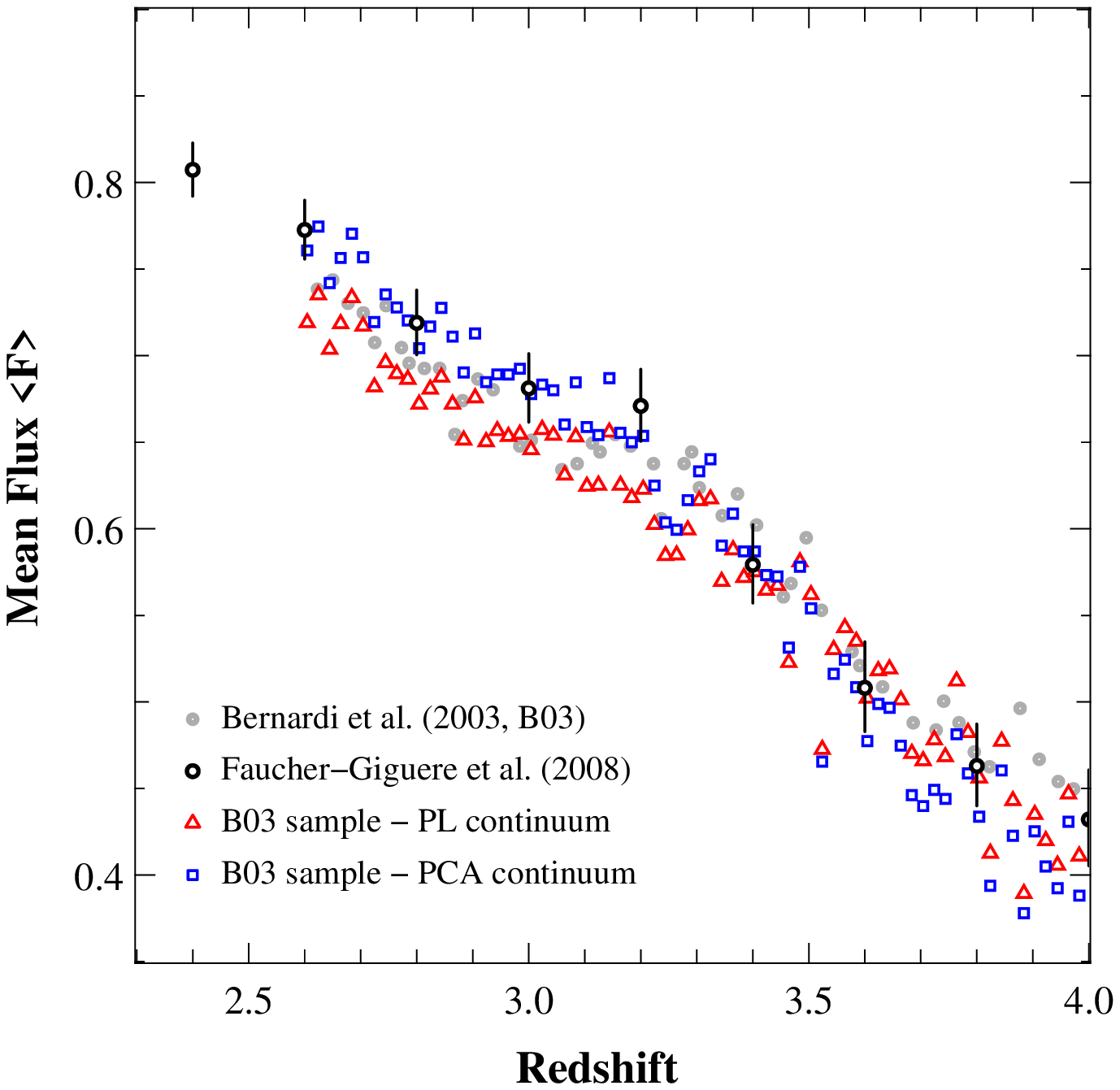}
\end{minipage}
\begin{minipage}[h]{.5\linewidth}
	\includegraphics[width=\linewidth]{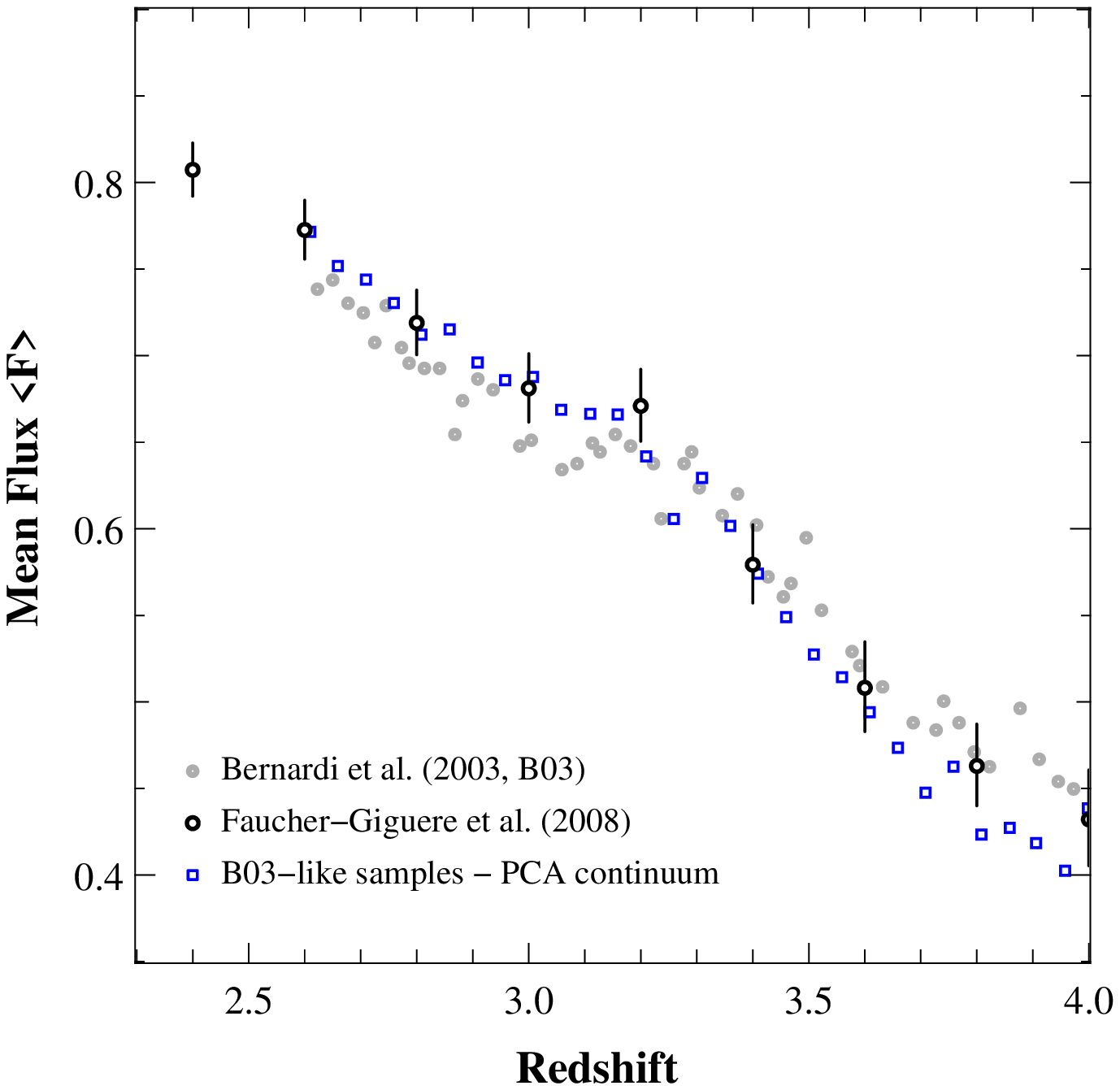}
\end{minipage}
\caption{ {\sl Left hand-side panel}: Mean flux redshift evolution inferred from power-law (red triangles) and 
PCA (blue squares) estimates of the continuum using a sample similar to \cite{bernardial2003}; the mean flux evolutions 
reported by \cite{bernardial2003} and \cite{FG2008} are shown as, respectively, grey points and
black open circles. Our measurements are consistent with B03 and FG08 results. In particular, a departure from a smooth evolution 
is seen at $z \sim 3.2$.
{\sl Right hand-side panel}:  500 samples are randomly drawn from the SDSS-DR7 (similar in size and redshift distribution to the B03 sample) and the mean flux evolution for each sample is then computed. The average of these measurements is displayed (PCA continuum, blue squares) and is in excellent agreement with \cite{FG2008} measurement.}
\label{fig:meanflux-B03}
\end{figure*}

We first would like to check if we can recover the \cite{bernardial2003} results.
The sample used in the \cite{bernardial2003} study is a sub-sample of SDSS-DR7 containing all the spectra observed 
up to the end of 2001 (corresponding to a modified Julian day $mjd = 52274$). Some selection
is applied to remove most prominent BALs and DLAs. These objects are not clearly defined in \cite{bernardial2003}
so that we have to apply our own selection. We avoid all BALs and DLAs as defined by Noterdaeme et al. (2009).
Our final sample has 837 QSOs when B03 had 1041.
The comparison to B03 is shown in \Fig{meanflux-B03} (left hand-side panel) for power-law (triangles) and PCA (squares) estimates of the 
continuum. Shown in the figure as well are the \cite{FG2008} results. As expected (see next Section), the power-law estimate is 
lower at $z<3$ than other estimates.
The evolutions found by us and B03 are in agreement and a departure from a smooth evolution is seen at $z \sim 3.2$.

We have randomly drawn from SDSS-DR7 a large number (500) of samples identical in size and redshift distribution
to the B03 sample. For each sample, we derived the mean flux observed at each redshift and calculated
the mean over the 500 samples.
The result is shown in \Fig{meanflux-B03} (right hand-side panel). 
The feature at $z\sim 3.2$ is still seen and could be a "bump" or a break in the evolution. 
Note that, as emphasized by \cite{FG2008}, a bump is seen only
if one insists on fitting a single power-law.

	\subsection{Redshift evolution of the mean flux in the IGM using SDSS-DR7}
	\label{s:DR7mf}
		
Continua of SDSS-DR7 spectra have been fitted using the two different methods we have described previously.
We derive a power-law and a PCA (using our and S05 sets of eigenspectra) continua. 
We restrict our study to spectra with a signal-to-noise ratio greater than 8 around 1280 \AA\ in the restframe 
to avoid instabilities in the fit of the power-law. We restrict the analysis to 
quasars with redshifts larger than $z>2.45$ to avoid the blue end of the spectra. 

Lines of sight containing damped \lya\ systems (DLAs) have been removed following the lists provided by \cite{pasquieral09} 
and \cite{Prochaskaal2005}. 
Broad absorption line quasars (BALs) flagged by \cite{shen10} have been avoided as well.
After this selection, we are left with 2,576
 quasars. Following \cite{bernardial2003}, we compute the mean flux
from the Lyman-$\alpha$ forest between 1080 and 1160~\AA\ in the restframe to
avoid the O~{\sc vi}-Lyman-$\beta$ and Lyman-$\alpha$ emission lines.\\

\begin{figure}
	\includegraphics[width=\linewidth]{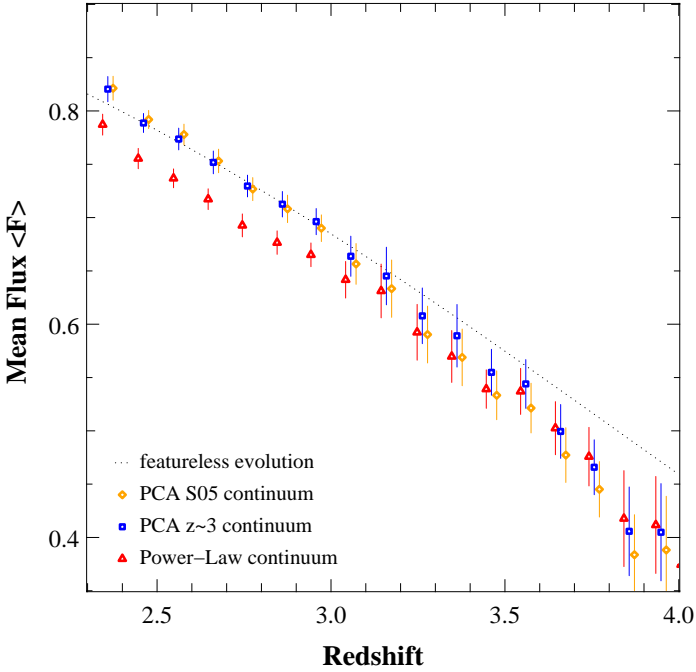}
\caption{Redshift evolution of the mean flux from SDSS DR7 quasar spectra. 
Three different estimates of the continuum are assumed (\Sec{cont-est}):
(i) an extrapolation of a power-law  fit (see \Sec{zqso}; red triangles); or a prediction (\Eq{eq_reconstruction}) using 
the output of a Principal Components Analysis of (ii) SDSS spectra at $z \sim 3$ as described in 
\Sec{pcaz3} (blue squares) or of (iii) HST spectra at $z \leq 1$ (S05; orange diamonds). 
Error bars are computed from bootstrapping and are at the 3$\sigma$ level. 
A change in the evolution can be noticed at $z\sim 3$ in the sense of a steeper slope at high redshift.
A featureless evolution (fitted from $z\leq 3$ points using $z\sim 3$ PCA  measurement) is shown for guidance.
}
\label{fig:meanflux}
\end{figure}

The mean flux in the \lya\  forest is then computed using three different continua: two PCA continua obtained from 
S05 principal components and the principal components derived in this work, and a power-law continuum. \Fig{meanflux} 
shows the redshift evolution of this quantity in redshift bins of size $\Delta z =0.1$. 
Error bars are computed from a bootstrap resampling.
There is apparently no difference in the results when using the two PCA decompositions derived
at low ($z\sim 1$) and high ($z\sim 3$) redshifts. 
The mean flux derived from the power law continuum is systematically smaller at $z<3$. This is 
expected as, the power-law tends to overestimate the continuum in the forest
(see \Sec{detectability}).

It is apparent that there is a change in the evolution of the mean flux at $z\sim 3$
with a steepening of the evolution at large redshift.

To estimate what kind of feature we are able to recover with our procedure,
we construct in the following DR7-like samples of simulated intermediate resolution 
quasar spectra probing an IGM with a mean flux evolution as seen in the high resolution data and test if we 
can recover this evolution by applying our procedures.

\subsection{Detectability of the bump with DR7}
\label{s:detectability}

In order to test the detectability of a feature equivalent to the bump seen at $z\sim 3.2$ by \cite{FG2008}, 
we performed simulations of this effect and its measurement with mock spectra.

	\subsubsection{Mock spectra}
		\label{sec:mockCreation}

We want to simulate the Lyman-$\alpha$ forest that will give a mean flux redshift evolution
similar to what is seen in high resolution data. This evolution was fitted by \cite{FG2008} as:
\begin{equation}
F(z) = e^{-A(1+z)^B-Ce^{{-\frac{[(1+z)-D]^2}{2E^2}}}},
\label{eq:faucher}
\end{equation}
where $A = 0.00153$, $B = 4.060$, $C = -0.0969$, $D = 4.267$ and $E = 0.0769$.\\

We assume that the Lyman-$\alpha$ forest is made up of absorption lines with a column density distribution 
\begin{equation}
f_{\rm NHI} = \alpha N_{\rm HI} ^{-\beta},
\label{eq:fnhi}
\end{equation}
with $\alpha = 4.9 \times 10^7$ and $\beta = 1.46$ over the column density range $10^{13} - 10^{17}$~cm$^{-2}$ and
a Doppler parameter distribution given by:
\begin{equation}
\frac{{\rm d}n}{{\rm d}b} = K \frac{b_{\sigma}^4}{b^5} e^{ -\frac{b_{\sigma }^4}{b^4} },
\label{eq:distriB}
\end{equation}
with $K = 6.82$ and $b_{\sigma} = 24.09\  \kms $ \citep{kim2001}.
Therefore the evolution in $F(z)$ is supposed to be due to an evolution in the number of clouds per unit redshift.
This is probably oversimplistic because, if any, the feature at $z\sim 3.2$ is claimed to be possibly due to 
an ionization process but this is probably fine for what we want to estimate.\\

We first construct the relation which gives at a given redshift the mean flux versus the number of cloud
present in a redshift bin of $\Delta z = 0.1$. This relation is parametrized by:
\begin{equation}
F_{\rm z}(n_{\rm abs}) = a_z e^{-\epsilon_{\rm z} n_{\rm abs}},
\label{eq:simul}
\end{equation}
where $n_{\rm abs}$ is the number of clouds drawn at random from the above population of clouds and
$a_{\rm z}$ and  $\epsilon_{\rm z}$ are determined by the simulation and depends on redshift.

Combining Eqs.~\ref{eq:faucher} and \ref{eq:simul}, we then derive at each redshift the actual number of clouds that is needed to reproduce the
relation given by \cite{FG2008}. We finally fit the redshift evolution of the number of clouds by a similar function:
\begin{equation}
\frac{dn}{dz} = n_0 (1+z)^{\gamma} + Ce^{{-\frac{[(1+z)-D]^2}{2E^2}}}.
\end{equation}
Our best fit gives the values $n_0 = 8.281 \pm 0.080$ and $\gamma = 3.076 \pm 0.006$, $C = -121.3 \pm 4.5$, $D = 4.267 \pm 0.003$ and $E = -0.081 \pm 0.003$ .\\

Once the number of clouds per unit redshift is correctly calibrated, we generate 50,000 mock spectra with a uniform 
emission redshift distribution in the range 2.3-4.5. For a given emission redshift, the number of absorption lines is computed from 
the line number density and is modulated to introduce Poisson noise. 
For each absorption line, the column density $N_{\rm HI}$ and the Doppler parameter are randomly chosen following \Eq{fnhi} and \Eq{distriB}. 
The spectrum is then degraded at the SDSS resolution ($R \sim 1800$) and a PCA continuum is added using 10 principal components and
choosing the weights at random within the calculated distributions. The wavelength scale is binned as for
SDSS spectra and noise is added following the SDSS g-magnitude distribution.

To check the validity of our procedures, 
the mean flux evolution is computed from spectra with no noise and no continuum added (see \Fig{detectabilityPlot}).
The mean flux evolution recovered by our procedure (grey diamonds) is in excellent agreement with the theoritical input 
(black dashed line) assumed to follow the evolution as derived by \cite{FG2008}.

\subsubsection{Should we detect any feature with DR7?}
\label{s:DetectFeature}

\begin{figure}
	\includegraphics[width=\linewidth]{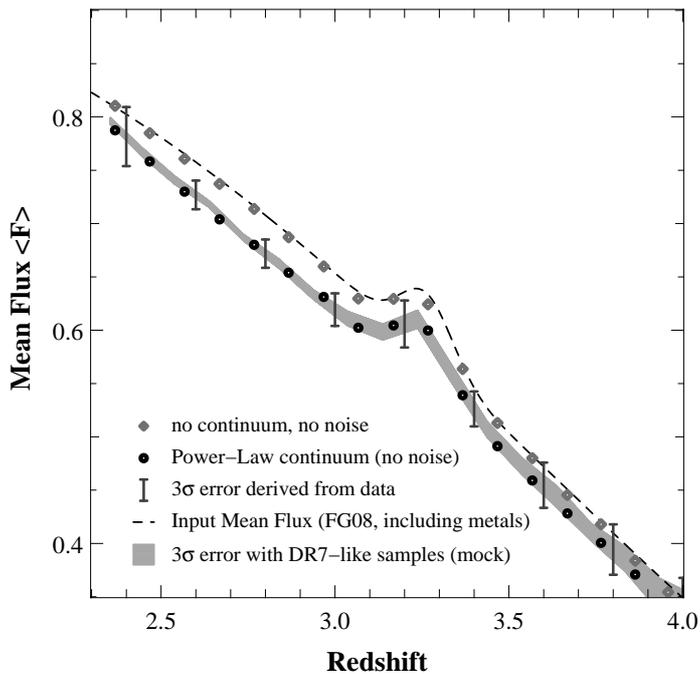}
\caption{Redshift evolution of the mean flux as measured from power-law continuum fitting of mock quasar spectra 
(black points and grey area). The evolution of the mean flux assumed as an input of the simulation is taken from \cite{FG2008} (black dashed line).
As already mentioned, the mean flux is underestimated by the power-law procedure but the shape of the evolution 
is recovered.
Vertical grey bars are the errors expected from the data. They are
computed from the errrors derived in the B03 sample (see \Fig{meanflux-B03}), scaled with respect to the 
different number of spectra in the SDSS-DR7 and B03 samples. They are, as expected, larger than the errors derived
from the simulations (grey area).}
\label{fig:detectabilityPlot}
\end{figure}

We compute 100 mock samples with the same number of quasars as the SDSS-DR7 and the same distributions of emission redshift and
signal-to-noise ratio. For each sample, we compute the mean flux evolution fitting quasar continuum with 
a power-law as performed on real data. At each redshift, we compute the scatter in the mean flux derived from these samples.
The recovered evolution is shown as black points in \Fig{detectabilityPlot}, to be compared with the 
dashed line showing the input assumed for the simulation. As already mentioned, it is apparent that the power-law fit of the QSO 
continuum underestimates the mean flux. However, the bump at $z\sim 3.2$, introduced in the input, is recovered although slightly smoothed out 
by the procedure. The errors derived from the simulations are shown as a grey area. They have to be compared with errors expected from
the data. The latter are estimated using the errors obtained in SDSS-DR7 from B03-like samples (as in \Fig{meanflux-B03}) but
scaled by the square root of the ratio of the number of quasars in B03 and SDSS-DR7 samples. These errors are shown in \Fig{detectabilityPlot}
by vertical error bars. They are as expected larger than the errors from the simulations.
Mock spectra are indeed idealized and additional sources of uncertainty are present in the calibration of the data
and the consequences on the continuum fit of the somewhat odd shape of some quasars.

\subsubsection{Feature at $z\sim 3.2$}
\label{s:ExistenceFeature}

\begin{figure}
	\includegraphics[width=\linewidth]{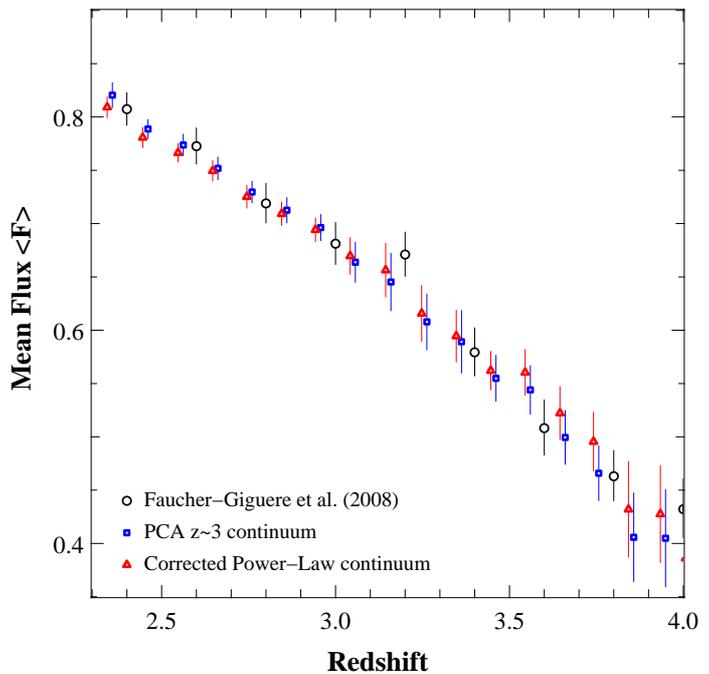}
\caption{Mean flux redshift evolution inferred from SDSS-DR7 quasars 
(red triangles: corrected power-law continuum, and blue squares: $z \sim 3$ PCA continuum) 
compared to the evolution from \cite{FG2008} (black open circles). The mean flux evolution derived with a power-law continuum 
is corrected from the bias predicted in simulation (see Fig.~17). All the measurements are in agreement with each other within errors. 
No "bump" is seen in any of the evolution 
inferred from SDSS-DR7 but there is a definite change in the slope of the evolution at $z\sim 3$ in the sense of a steeper
evolution beyond this redshift.}
\label{fig:meanflux-DR7}
\end{figure}

We summarize in \Fig{meanflux-DR7} the mean flux evolution measured from the SDSS-DR7 data
for a continuum estimated with a PCA (blue squares) or a power law (red triangles) corrected for the
systematic bias as seen in \Fig{detectabilityPlot}. 
Indeed, the power-law continuum systematically overestimates the amount of absorption in the \lya\ forest.
The mean flux evolution derived with a 
power-law continuum is corrected by the expected difference seen in mocks between the measured mean flux
and the input of the simulation.

There is an excellent agreement between the results of the two methods. Overplotted as black circles 
on \Fig{meanflux-DR7} are the results by \cite{FG2008} in very good agreement with  
the PCA estimate except, may be, in the bin
around $z\sim 3.2$. Note that the discrepancy is less than 2$\sigma$. 
In any case, it is apparent that the smooth redshift evolution of the mean flux becomes steeper around redshift $z\sim 3$. 

The slight difference between high and intermediate resolution data at $z\sim 3.2$ may be explained by a different selection
of the quasars. Indeed, \cite{worseck10} have argued that SDSS preferentially selects 
3$<$$z_{\rm em}$$<$3.5 quasars with intervening H~{\sc i} Lyman limit systems. However, this should
have little influence on the mean flux in the overwhole forest and cannot explain the discrepancy
by itself. This could also be due to the possibility that our procedure smoothes out a sharp feature. 
However, this would be surprising given the width of the bins and the results of our simulations
(see \Fig{detectabilityPlot}).

\section{Conclusion}
\label{s:Conclu}
The first goal of this paper is to provide a new PCA decomposition of the continuum of
$z\sim 3$ quasars. This should be useful for studies of the Lyman-$\alpha$ forest
and to generate mock quasar continua to be implemented in simulations constructed to search for systematic effects 
in future analysis or surveys.
We took the opportunity to enlarge to $1020-2000$~\AA~ the wavelength range over which the spectra
are decomposed, to be compared with the previous wavelength range $1020-1600$~\AA~ used by S05.
The mean spectrum at $z\sim 3$ has a similar shape as the mean spectrum derived at $z\sim 1$ by S05 except
that the strength of the Lyman-$\alpha$ and C~{\sc iv} emission lines relative to the continuum is smaller 
at high redshift. Our work concentrates on the estimate of the continuum in the Lyman-$\alpha$ forest and
we provide all outputs of this analysis that are required to generate mock continua.

We use this decomposition to revisit the evolution with redshift of the mean flux
in the Lyman-$\alpha$ forest and to compare two methods to estimate the quasar flux in the Lyman-$\alpha$ forest:
the extrapolation of a power-law fitted to the red part of the spectrum and an estimate of the flux
from PCA coefficients. \\
We find that:\\
(i) the power law method systematically underestimates the mean flux by an amount decreasing with redshift. When
correcting for this bias, as estimated with simulations, we find that the method gives similar results as the PCA method;\\
(ii) the PCA method yields results very similar to what is measured by Bernardi et al. (2003) and from 
high spectral resolution data \citep{FG2008};\\
(iii) from our simulations a bump at $z\sim 3.2$, if present, should be marginally detected with the data set 
of SDSS-DR7;\\  
(iv) finally, from our analysis, we find that there is a definite
break in the evolution of the mean flux at $z\sim 3$ in the sense of a steeper decrease of the mean flux at high redshift.
We caution that this could be a consequence of a more prominent bump to be slightly smoothed out by our procedures.

The increase of the statistics but most importantly of the quality of the data
that will be soon provided by the BOSS survey should definitely settle this issue.

\begin{acknowledgements}
We thank an anonymous referee for important and very useful comments.
This project was  supported by the Agence Nationale de la Recherche under contract ANR-08-BLAN-0222.
\end{acknowledgements}

\bibliographystyle{aa}
\bibliography{PCA} 

\begin{thebibliography}{54}
\expandafter\ifx\csname natexlab\endcsname\relax\def\natexlab#1{#1}\fi

\bibitem[{{Baldwin}(1977)}]{baldwin1977}
{Baldwin}, J.~A. 1977, \apj, 214, 679

\bibitem[{{Becker} {et~al.}(2007){Becker}, {Rauch}, \& {Sargent}}]{Becker07}
{Becker}, G.~D., {Rauch}, M., \& {Sargent}, W.~L.~W. 2007, \apj, 662, 72

\bibitem[{{Bergeron} {et~al.}(2004){Bergeron}, {Petitjean}, {Aracil}, {Pichon},
  {Scannapieco}, {Srianand}, {Boisse}, {Carswell}, {Chand}, {Cristiani},
  {Ferrara}, {Haehnelt}, {Hughes}, {Kim}, {Ledoux}, {Richter}, \&
  {Viel}}]{Bergeron04}
{Bergeron}, J., {Petitjean}, P., {Aracil}, B., {et~al.} 2004, The Messenger,
  118, 40

\bibitem[{{Bernardi} {et~al.}(2003){Bernardi}, {Sheth}, {SubbaRao}, {Richards},
  {Burles}, {Connolly}, {Frieman}, {Nichol}, {Schaye}, {Schneider}, {Vanden
  Berk}, {York}, {Brinkmann}, \& {Lamb}}]{bernardial2003}
{Bernardi}, M., {Sheth}, R.~K., {SubbaRao}, M., {et~al.} 2003, \aj, 125, 32

\bibitem[{{Bi} {et~al.}(1992){Bi}, {Boerner}, \& {Chu}}]{Bi92}
{Bi}, H.~G., {Boerner}, G., \& {Chu}, Y. 1992, \aap, 266, 1

\bibitem[{{Bolton} \& {Haehnelt}(2007)}]{Boltonal07}
{Bolton}, J.~S. \& {Haehnelt}, M.~G. 2007, \mnras, 382, 325

\bibitem[{{Bolton} {et~al.}(2005){Bolton}, {Haehnelt}, {Viel}, \&
  {Springel}}]{Bolton05}
{Bolton}, J.~S., {Haehnelt}, M.~G., {Viel}, M., \& {Springel}, V. 2005, \mnras,
  357, 1178

\bibitem[{{Cen} {et~al.}(1994){Cen}, {Miralda-Escud{\'e}}, {Ostriker}, \&
  {Rauch}}]{Cen94}
{Cen}, R., {Miralda-Escud{\'e}}, J., {Ostriker}, J.~P., \& {Rauch}, M. 1994,
  \apjl, 437, L9

\bibitem[{{Croft} {et~al.}(1998){Croft}, {Weinberg}, {Katz}, \&
  {Hernquist}}]{Croft98}
{Croft}, R.~A.~C., {Weinberg}, D.~H., {Katz}, N., \& {Hernquist}, L. 1998,
  \apj, 495, 44

\bibitem[{{Dall'Aglio} {et~al.}(2008){Dall'Aglio}, {Wisotzki}, \&
  {Worseck}}]{Dallaglioal08}
{Dall'Aglio}, A., {Wisotzki}, L., \& {Worseck}, G. 2008, \aap, 491, 465

\bibitem[{{Dall'Aglio} {et~al.}(2009){Dall'Aglio}, {Wisotzki}, \&
  {Worseck}}]{Dallaglioal09}
{Dall'Aglio}, A., {Wisotzki}, L., \& {Worseck}, G. 2009,
  ArXiv:astro-ph/0906.1484

\bibitem[{{Eisenstein} {et~al.}(2011){Eisenstein}, {Weinberg}, {Agol},
  {Aihara}, {Allende Prieto}, {Anderson}, {Arns}, {Aubourg}, {Bailey},
  {Balbinot}, \& et~al.}]{Eisenstein11}
{Eisenstein}, D.~J., {Weinberg}, D.~H., {Agol}, E., {et~al.} 2011,
  ArXiv:astro-ph/1101.1529

\bibitem[{{Fan}(2009)}]{fan2009}
{Fan}, X. 2009, in Astronomical Society of the Pacific Conference Series, Vol.
  408, Astronomical Society of the Pacific Conference Series, ed. {W.~Wang,
  Z.~Yang, Z.~Luo, \& Z.~Chen}, 439

\bibitem[{{Fan} {et~al.}(2006){Fan}, {Carilli}, \& {Keating}}]{Fan06}
{Fan}, X., {Carilli}, C.~L., \& {Keating}, B. 2006, \araa, 44, 415

\bibitem[{{Faucher-Gigu{\`e}re} {et~al.}(2008){Faucher-Gigu{\`e}re},
  {Prochaska}, {Lidz}, {Hernquist}, \& {Zaldarriaga}}]{FG2008}
{Faucher-Gigu{\`e}re}, C.-A., {Prochaska}, J.~X., {Lidz}, A., {Hernquist}, L.,
  \& {Zaldarriaga}, M. 2008, \apj, 681, 831

\bibitem[{{Francis} {et~al.}(1992){Francis}, {Hewett}, {Foltz}, \&
  {Chaffee}}]{francis92}
{Francis}, P.~J., {Hewett}, P.~C., {Foltz}, C.~B., \& {Chaffee}, F.~H. 1992,
  \apj, 398, 476

\bibitem[{{Francis} {et~al.}(1991){Francis}, {Hewett}, {Foltz}, {Chaffee},
  {Weymann}, \& {Morris}}]{francis91}
{Francis}, P.~J., {Hewett}, P.~C., {Foltz}, C.~B., {et~al.} 1991, \apj, 373,
  465

\bibitem[{{Gunn} \& {Peterson}(1965)}]{GunnandPeterson}
{Gunn}, J.~E. \& {Peterson}, B.~A. 1965, \apj, 142, 1633

\bibitem[{{Hennawi} \& {Prochaska}(2007)}]{hennawi2007}
{Hennawi}, J.~F. \& {Prochaska}, J.~X. 2007, \apj, 655, 735

\bibitem[{{Hernquist} {et~al.}(1996){Hernquist}, {Katz}, {Weinberg}, \&
  {Miralda-Escud{\'e}}}]{Hernquist96}
{Hernquist}, L., {Katz}, N., {Weinberg}, D.~H., \& {Miralda-Escud{\'e}}, J.
  1996, \apjl, 457, L51

\bibitem[{{Hu} {et~al.}(1995){Hu}, {Kim}, {Cowie}, {Songaila}, \&
  {Rauch}}]{Hu95}
{Hu}, E.~M., {Kim}, T.-S., {Cowie}, L.~L., {Songaila}, A., \& {Rauch}, M. 1995,
  \aj, 110, 1526

\bibitem[{{Kim} {et~al.}(2001){Kim}, {Cristiani}, \& {D'Odorico}}]{kim2001}
{Kim}, T., {Cristiani}, S., \& {D'Odorico}, S. 2001, \aap, 373, 757

\bibitem[{{Kim} {et~al.}(2007){Kim}, {Bolton}, {Viel}, {Haehnelt}, \&
  {Carswell}}]{Kim07}
{Kim}, T.-S., {Bolton}, J.~S., {Viel}, M., {Haehnelt}, M.~G., \& {Carswell},
  R.~F. 2007, \mnras, 382, 1657

\bibitem[{{Kirkman} {et~al.}(2005){Kirkman}, {Tytler}, {Suzuki}, {Melis},
  {Hollywood}, {James}, {So}, {Lubin}, {Jena}, {Norman}, \&
  {Paschos}}]{Kirkman2005}
{Kirkman}, D., {Tytler}, D., {Suzuki}, N., {et~al.} 2005, \mnras, 360, 1373

\bibitem[{{Lynds}(1971)}]{Lynds}
{Lynds}, R. 1971, \apjl, 164, L73

\bibitem[{{McDonald} {et~al.}(2005){McDonald}, {Seljak}, {Cen}, {Shih},
  {Weinberg}, {Burles}, {Schneider}, {Schlegel}, {Bahcall}, {Briggs},
  {Brinkmann}, {Fukugita}, {Ivezi{\'c}}, {Kent}, \& {Vanden
  Berk}}]{McDonaldal05a}
{McDonald}, P., {Seljak}, U., {Cen}, R., {et~al.} 2005, \apj, 635, 761

\bibitem[{{Noterdaeme} {et~al.}(2009){Noterdaeme}, {Petitjean}, {Ledoux}, \&
  {Srianand}}]{pasquieral09}
{Noterdaeme}, P., {Petitjean}, P., {Ledoux}, C., \& {Srianand}, R. 2009, \aap,
  505, 1087

\bibitem[{{Oke} \& {Korycansky}(1982)}]{Okeal82}
{Oke}, J.~B. \& {Korycansky}, D.~G. 1982, \apj, 255, 11

\bibitem[{{Petitjean} {et~al.}(1995){Petitjean}, {Mueket}, \&
  {Kates}}]{petitjean95}
{Petitjean}, P., {Mueket}, J.~P., \& {Kates}, R.~E. 1995, \aap, 295, L9

\bibitem[{{Petitjean} {et~al.}(1998){Petitjean}, {Surdej}, {Smette}, {Shaver},
  {M\"uecket}, \& {Remy}}]{petitjean98}
{Petitjean}, P., {Surdej}, J., {Smette}, A., {et~al.} 1998, \aap, 334, L45

\bibitem[{{Petitjean} {et~al.}(1993){Petitjean}, {Webb}, {Rauch}, {Carswell},
  \& {Lanzetta}}]{petitjean93}
{Petitjean}, P., {Webb}, J.~K., {Rauch}, M., {Carswell}, R.~F., \& {Lanzetta},
  K. 1993, \mnras, 262, 499

\bibitem[{{Press} {et~al.}(1992){Press}, {Teukolsky}, {Vetterling}, \&
  {Flannery}}]{press92}
{Press}, W.~H., {Teukolsky}, S.~A., {Vetterling}, W.~T., \& {Flannery}, B.~P.
  1992, {Numerical recipes in C. The art of scientific computing}, ed. {Press,
  W.~H., Teukolsky, S.~A., Vetterling, W.~T., \& Flannery, B.~P. } (Cambridge:
  University Press)

\bibitem[{{Prochaska} {et~al.}(2005){Prochaska}, {Herbert-Fort}, \&
  {Wolfe}}]{Prochaskaal2005}
{Prochaska}, J.~X., {Herbert-Fort}, S., \& {Wolfe}, A.~M. 2005, \apj, 635, 123

\bibitem[{{Prochaska} {et~al.}(2009){Prochaska}, {Worseck}, \&
  {O'Meara}}]{Prochaskaal09}
{Prochaska}, J.~X., {Worseck}, G., \& {O'Meara}, J.~M. 2009, \apjl, 705, L113

\bibitem[{{Rauch}(1998)}]{Rauch98}
{Rauch}, M. 1998, \araa, 36, 267

\bibitem[{{Rauch} {et~al.}(1997){Rauch}, {Miralda-Escude}, {Sargent}, {Barlow},
  {Weinberg}, {Hernquist}, {Katz}, {Cen}, \& {Ostriker}}]{Rauchal97}
{Rauch}, M., {Miralda-Escude}, J., {Sargent}, W.~L.~W., {et~al.} 1997, \apj,
  489, 7

\bibitem[{{Riediger} {et~al.}(1998){Riediger}, {Petitjean}, \&
  {M\"ucket}}]{riediger98}
{Riediger}, R., {Petitjean}, P., \& {M\"ucket}, J.~P. 1998, \aap, 329, 30

\bibitem[{{Schaye} {et~al.}(2000){Schaye}, {Theuns}, {Rauch}, {Efstathiou}, \&
  {Sargent}}]{Schaye00}
{Schaye}, J., {Theuns}, T., {Rauch}, M., {Efstathiou}, G., \& {Sargent},
  W.~L.~W. 2000, \mnras, 318, 817

\bibitem[{{Schlegel} {et~al.}(2009){Schlegel}, {Bebek}, {Heetderks}, {Ho},
  {Lampton}, {Levi}, {Mostek}, {Padmanabhan}, {Perlmutter}, {Roe}, {Sholl},
  {Smoot}, {White}, {Dey}, {Abraham}, {Jannuzi}, {Joyce}, {Liang}, {Merrill},
  {Olsen}, \& {Salim}}]{BigBOSS}
{Schlegel}, D.~J., {Bebek}, C., {Heetderks}, H., {et~al.} 2009,
  ArXiv:astro-ph/0904.0468

\bibitem[{{Schlegel} {et~al.}(2007){Schlegel}, {Blanton}, {Eisenstein},
  {Gillespie}, {Gunn}, {Harding}, {McDonald}, {Nichol}, {Padmanabhan},
  {Percival}, {Richards}, {Rockosi}, {Roe}, {Ross}, {Schneider}, {Strauss},
  {Weinberg}, \& {White}}]{Schlegelal07}
{Schlegel}, D.~J., {Blanton}, M., {Eisenstein}, D., {et~al.} 2007, in Bulletin
  of the American Astronomical Society, Vol.~38, 966

\bibitem[{{Shen} {et~al.}(2010){Shen}, {Hall}, {Richards}, {Schneider},
  {Strauss}, {Snedden}, {Bizyaev}, {Brewington}, {Malanushenko},
  {Malanushenko}, {Oravetz}, {Pan}, \& {Simmons}}]{shen10}
{Shen}, Y., {Hall}, P.~B., {Richards}, G.~T., {et~al.} 2010,
  ArXiv:astro-ph/1006.5178

\bibitem[{{Slosar} {et~al.}(2009){Slosar}, {Ho}, {White}, \&
  {Louis}}]{slosaral09}
{Slosar}, A., {Ho}, S., {White}, M., \& {Louis}, T. 2009, Journal of Cosmology
  and Astro-Particle Physics, 10, 19

\bibitem[{{Songaila}(2004)}]{songaila2004}
{Songaila}, A. 2004, \aj, 127, 2598

\bibitem[{{Suzuki} {et~al.}(2005){Suzuki}, {Tytler}, {Kirkman}, {O'Meara}, \&
  {Lubin}}]{Suzukial05}
{Suzuki}, N., {Tytler}, D., {Kirkman}, D., {O'Meara}, J.~M., \& {Lubin}, D.
  2005, \apj, 618, 592

\bibitem[{{Theuns} {et~al.}(2002){Theuns}, {Bernardi}, {Frieman}, {Hewett},
  {Schaye}, {Sheth}, \& {Subbarao}}]{Theuns02c}
{Theuns}, T., {Bernardi}, M., {Frieman}, J., {et~al.} 2002, \apjl, 574, L111

\bibitem[{{Theuns} {et~al.}(1998){Theuns}, {Leonard}, {Efstathiou}, {Pearce},
  \& {Thomas}}]{Theuns98}
{Theuns}, T., {Leonard}, A., {Efstathiou}, G., {Pearce}, F.~R., \& {Thomas},
  P.~A. 1998, \mnras, 301, 478

\bibitem[{{Theuns} \& {Srianand}(2006)}]{theunsal06}
{Theuns}, T. \& {Srianand}, R. 2006, in IAU Symposium, Vol. 232, The Scientific
  Requirements for Extremely Large Telescopes, ed. {P.~Whitelock, M.~Dennefeld,
  \& B.~Leibundgut}, 464--471

\bibitem[{{Tytler} {et~al.}(2004){Tytler}, {Kirkman}, {O'Meara}, {Suzuki},
  {Orin}, {Lubin}, {Paschos}, {Jena}, {Lin}, {Norman}, \&
  {Meiksin}}]{Tytleral04}
{Tytler}, D., {Kirkman}, D., {O'Meara}, J.~M., {et~al.} 2004, \apj, 617, 1

\bibitem[{{Vanden Berk} {et~al.}(2001){Vanden Berk}, {Richards}, {Bauer},
  {Strauss}, {Schneider}, {Heckman}, {York}, {Hall}, {Fan}, {Knapp},
  {Anderson}, {Annis}, {Bahcall}, {Bernardi}, {Briggs}, {Brinkmann}, {Brunner},
  {Burles}, {Carey}, {Castander}, {Connolly}, {Crocker}, {Csabai}, {Doi},
  {Finkbeiner}, {Friedman}, {Frieman}, {Fukugita}, {Gunn}, {Hennessy},
  {Ivezi{\'c}}, {Kent}, {Kunszt}, {Lamb}, {Leger}, {Long}, {Loveday}, {Lupton},
  {Meiksin}, {Merelli}, {Munn}, {Newberg}, {Newcomb}, {Nichol}, {Owen}, {Pier},
  {Pope}, {Rockosi}, {Schlegel}, {Siegmund}, {Smee}, {Snir}, {Stoughton},
  {Stubbs}, {SubbaRao}, {Szalay}, {Szokoly}, {Tremonti}, {Uomoto}, {Waddell},
  {Yanny}, \& {Zheng}}]{VB01}
{Vanden Berk}, D.~E., {Richards}, G.~T., {Bauer}, A., {et~al.} 2001, \aj, 122,
  549

\bibitem[{{White} {et~al.}(2010){White}, {Pope}, {Carlson}, {Heitmann},
  {Habib}, {Fasel}, {Daniel}, \& {Lukic}}]{whiteal09}
{White}, M., {Pope}, A., {Carlson}, J., {et~al.} 2010, \apj, 713, 383

\bibitem[{{Worseck} \& {Prochaska}(2011)}]{worseck10}
{Worseck}, G. \& {Prochaska}, J.~X. 2011, \apj, 728, 23

\bibitem[{{Yip} {et~al.}(2004){Yip}, {Connolly}, {Vanden Berk}, {Ma},
  {Frieman}, {SubbaRao}, {Szalay}, {Richards}, {Hall}, {Schneider}, {Hopkins},
  {Trump}, \& {Brinkmann}}]{Yipal04}
{Yip}, C.~W., {Connolly}, A.~J., {Vanden Berk}, D.~E., {et~al.} 2004, \aj, 128,
  2603

\bibitem[{{Zhang} {et~al.}(1995){Zhang}, {Anninos}, \& {Norman}}]{Zhang95}
{Zhang}, Y., {Anninos}, P., \& {Norman}, M.~L. 1995, \apjl, 453, L57

\bibitem[{{Zheng} {et~al.}(1997){Zheng}, {Kriss}, {Telfer}, {Grimes}, \&
  {Davidsen}}]{zheng97}
{Zheng}, W., {Kriss}, G.~A., {Telfer}, R.~C., {Grimes}, J.~P., \& {Davidsen},
  A.~F. 1997, \apj, 475, 469

\end{thebibliography}
\end{document}